\documentclass[fleqn,usenatbib,useAMS]{mnras}

\usepackage{newtxtext}

\usepackage{float}
\usepackage{graphicx}
\usepackage{times}
\usepackage{amstext}
\usepackage{amsmath}
\usepackage{amssymb}	
\usepackage{natbib}
\usepackage{url}
\usepackage{tabularx}
\usepackage{color}
\usepackage{graphics}
\usepackage{multirow}
\usepackage{xspace}
\mathchardef\mhyphen="2D
\newcommand{\kms}{\ensuremath{\mathrm{\,km\,s}^{-1}}\xspace} 

\newcommand{\solar}{M$_{\odot}$}
\newcommand{\farc}{\overset{\prime\prime}{.}}
\newcommand{\fdeg}{\overset{\circ}{.}}
\newcommand{\brick}{G0.253+0.016}

\usepackage[T1]{fontenc}
\usepackage{ae,aecompl}

\usepackage{txfonts}

\DeclareRobustCommand{\VAN}[3]{#2}
\let\VANthebibliography\thebibliography
\def\thebibliography{\DeclareRobustCommand{\VAN}[3]{##3}\VANthebibliography}

\definecolor{darkgreen}{rgb}{0.0, 0.5, 0.0}

\title[A bubble in the Brick]{A wind-blown bubble in the Central Molecular Zone cloud G0.253+0.016}

\author[J. D. Henshaw et al.]{Jonathan D. Henshaw,$^{1}$\thanks{E-mail: jonathan.d.henshaw@gmail.com}
Mark~R.~Krumholz,$^{1,2,3,4}$
Natalie~O.~Butterfield,$^{5}$
Jonathan~Mackey,$^{6,7}$
\newauthor 
Adam~Ginsburg,$^{8}$
Thomas~J.~Haworth,$^{9}$
Francisco~Nogueras-Lara,$^{1}$
Ashley~T.~Barnes,$^{10}$
Steven~N.~Longmore,$^{11}$
\newauthor
John~Bally,$^{12}$
J.~M.~Diederik~Kruijssen,$^{13}$
Elisabeth~A.~C.~Mills,$^{14}$
Henrik~Beuther,$^{1}$
Daniel~L.~Walker,$^{15}$
\newauthor
Cara~Battersby,$^{15}$
Alyssa~Bulatek,$^{8}$
Thomas~Henning,$^{1}$
Juergen~Ott,$^{16, 17}$
and
Juan~D.~Soler,$^{1}$
\\
$^{1}$ Max Planck Institute for Astronomy, K\"{o}nigstuhl 17, D-69117 Heidelberg, Germany\\
$^{2}$ Research School of Astronomy and Astrophysics, Australian National University, Canberra, ACT 2611 Australia\\
$^{3}$ ARC Centre of Excellence for Astronomy in Three Dimensions (ASTRO-3D), Canberra, ACT 2611 Australia\\
$^4$ Institut fur Theoretische Astrophysik, Zentrum f\"ur Astronomie, Universit\"at Heidelberg, D-69120 Heidelberg, Germany\\
$^{5}$ Department of Physics, Villanova University, 800 E. Lancaster Ave.,
Villanova, PA 19085, USA\\
$^{6}$ Dublin Institute for Advanced Studies, Astronomy \& Astrophysics Section, 31 Fitzwilliam Place, Dublin 2, Ireland\\
$^{7}$ Dublin Institute for Advanced Studies, Centre for AstroParticle Physics and Astrophysics (CAPPA), DIAS Dunsink Observatory, Dunsink Lane, Dublin 15, Ireland\\
$^{8}$ Department of Astronomy, University of Florida, PO Box 112055, USA \\
$^{9}$ Astronomy Unit, School of Physics and Astronomy, Queen Mary University of London, London E1 4NS, UK\\
$^{10}$ Argelander Institute f\"{u}r Astronomy, University of Bonn, Auf dem H\"{u}gel 71, 53121 Bonn, Germany\\
$^{11}$ Astrophysics Research Institute, Liverpool John Moores University, IC2, 146 Brownlow Hill, Liverpool, L3 5RF, United Kingdom\\
$^{12}$ CASA, University of Colorado, 389-UCB, Boulder, CO 80309 \\
$^{13}$ Astronomisches Rechen-Institut, Zentrum f\"{u}r Astronomie der Universit\"{a}t Heidelberg, M\"{o}nchhofstra{\ss}e 12-14, 69120 Heidelberg, Germany\\
$^{14}$ Department of Physics and Astronomy, University of Kansas, 1251 Wescoe Hall Drive, Lawrence, KS 66045, USA \\
$^{15}$ University of Connecticut, Department of Physics, 196A Auditorium Road, Unit 3046, Storrs, CT 06269 USA \\
$^{16}$ National Radio Astronomy Observatory, 1003 Lopezville Road, Socorro, NM 87801, USA \\
$^{17}$ New Mexico Institute of Mining and Technology, 801 Leroy Place, Socorro, NM 87801, USA
}

\date{Accepted 2021 October 15. Received 2021 October 13; in original form 2021 August 16.}

\pubyear{2021}

\begin{document}
\label{firstpage}
\pagerange{\pageref{firstpage}--\pageref{lastpage}}
\maketitle

\begin{abstract}
G0.253+0.016, commonly referred to as ``the Brick'' and located within the Central Molecular Zone, is one of the densest ($\approx10^{3-4}$\,cm$^{-3}$) molecular clouds in the Galaxy to lack signatures of widespread star formation. 
We set out to constrain the origins of an arc-shaped molecular line emission feature located within the cloud. We determine that the arc, centred on $\{l_{0},b_{0}\}=\{0\fdg248,\,0\fdg018\}$, has a radius of $1.3$\,pc and kinematics indicative of the presence of a shell expanding at $5.2^{+2.7}_{-1.9}$\,\kms. Extended radio continuum emission fills the arc cavity and recombination line emission peaks at a similar velocity to the arc, implying that the molecular and ionised gas are physically related. The inferred Lyman continuum photon rate is $N_{\rm LyC}=10^{46.0}\mhyphen10^{47.9}$\,photons\,s$^{-1}$, consistent with a star of spectral type B1-O8.5, corresponding to a mass of $\approx12\mhyphen20$\,\solar. We explore two scenarios for the origin of the arc: i) a partial shell swept up by the wind of an interloper high-mass star; ii) a partial shell swept up by stellar feedback resulting from in-situ star formation. We favour the latter scenario, finding reasonable (factor of a few) agreement between its morphology, dynamics, and energetics and those predicted for an expanding bubble driven by the wind from a high-mass star. The immediate implication is that \brick \ may not be as quiescent as is commonly accepted. We speculate that the cloud may have produced a $\lesssim10^{3}$\,\solar \ star cluster $\gtrsim0.4$\,Myr ago, and demonstrate that the high-extinction and stellar crowding observed towards \brick \ may help to obscure such a star cluster from detection.
\end{abstract}




\section{Introduction}

The Central Molecular Zone (hereafter, CMZ), i.e.\ the inner few hundred parsecs of the Milky Way, hosts some of the Galaxy's densest molecular clouds \citep{lis_1994b, bally_2010, longmore_2012,longmore_2013b,walker_2015, mills_2018} and star clusters \citep[known as the Arches and Quintuplet;][]{Figer1999, portegies-zwart_2010, longmore_2014}. Of the former, G0.253+0.016 (often referred to as ``the Brick'') is probably one of the most enigmatic molecular clouds in the Galaxy. Much of the interest in this cloud stems from the fact that it exhibits little evidence of widespread star formation activity \citep{lis_1994, immer_2012, mills_2015}, in spite of its high mass ($\approx10^{5}$\,\solar) and mean density \citep[$\approx10^{3-4}$\,cm$^{-3}$;][]{lis_1994b, lis_1998b, longmore_2012, rathborne_2014, mills_2018}. 

Until recently, the only direct evidence for star formation within \brick \ was a single water maser \citep[see also \citealp{Lu2019b}]{lis_1994}. This evidence has been strengthened considerably by recent high-angular resolution Atacama Large Millimeter/submillimeter Array (ALMA) observations of the maser source, which reveal a small cluster of low-to-intermediate mass protostars, 50\% of which are driving bi-polar outflows \citep{Walker2021}. Deep radio continuum observations and additional searches for maser emission have not revealed any further star formation activity \citep{immer_2012, mills_2015, rodriguez_2013, Lu2019}, and all other evidence for star formation comes from indirect energy balance arguments. \citet{lis_2001} model the far-infrared/sub-millimetre spectral energy distribution of \brick, and infer that the cloud's luminosity is conceivably generated by four B0 zero-age main-sequence stars. \citet{marsh_2016} report evidence of heated dust emission that follows a tadpole-shaped ridge, which they suggest may result from a chain of embedded protostars. 

Clouds with the physical characteristics of G0.253+0.016, but which are not already prodigiously forming stars, do not exist within the Milky Way disc \citep{ginsburg_2012,Urquhart2018}. Consequently, G0.253+0.016 presents a unique opportunity to study the early phases of high-mass star and cluster formation under the extreme conditions found in the Galactic Centre \citep{longmore_2012,longmore_2013b, rathborne_2014a}. Recent observational work has set out to categorise G0.253+0.016's internal structure and dynamics in order to better understand its star formation potential. The internal structure of the cloud is complex \citep{kauffmann_2013, henshaw_2019}. Dust continuum and molecular line observations reveal significant sub-structure, with a few dozen compact cores and filaments detected in both emission and absorption \citep{bally_2014, johnston_2014, rathborne_2014, rathborne_2015, federrath_2016, Battersby2020, Hatchfield2020}. Gas motions measured on $\sim0.1$\,pc scales are highly supersonic \citep{henshaw_2019,henshaw_2020}, resulting in widespread shocked gas emission \citep{kauffmann_2013, johnston_2014}.

\citet{federrath_2016} inferred that the internal turbulence in \brick \ is dominated by solenoidal motion, likely resulting from the strong shear induced by its eccentric orbit around the Galactic Centre \citep{kruijssen_2019}. The shear resulting from the background gravitational potential and the cloud's orbital motion may help to explain its morphology \citep{kruijssen_2019, Petkova2021}. The combination of solenoidal gas motion, a strong magnetic field \citep{Pillai2015}, and an elevated critical density threshold for star formation \citep{kruijssen_2014b, rathborne_2014, ginsburg_2018} may explain the overall low star formation rate of \brick.

However, there is a complication to this simple picture, in the form of an arcuate, shell-like structure detected within the cloud's interior. It has been detected in a variety of molecular species including SO \citep{higuchi_2014}, NH$_{3}$ \citep{mills_2015}, HNCO \citep{henshaw_2019}, and SiO \citep{Walker2021}. Both the gas and dust temperature along the rim of the arc appear to be elevated, evidenced by its clear detection in higher-excitation lines of NH$_{3}$ [\citealt{mills_2015} report detections in the (6,6) and (7,7) inversion transitions]. The arc is also co-spatial with the spine of warm dust identified by \citet{marsh_2016}. Class I Methanol masers, believed to be tracing shocked gas emission that is not directly related to star formation (unlike Class II masers), are furthermore detected in a crescent-like arrangement following the arc emission observed in NH$_{3}$ \citep{mills_2015}. Following detailed investigation of the dynamics of G0.253+0.016, \citet{henshaw_2019} demonstrated that the arc is coherent in both projected space and in velocity. The bulk of the emission associated with \brick \ is spread over a velocity range of $\sim40$\,\kms. In position-position-velocity space, there are at least two cloud components. The ``main'' component is that which closely resembles \brick \ as it appears in dust continuum emission, and has a mean velocity of $\sim37$\,\kms. The mean velocity of the component associated with the arc is $\sim17$\,\kms. However, the velocity gradient associated with this latter component is such that this and the main component appear to meet (in position-position-velocity space) towards the south of the cloud \citep{henshaw_2019}. 

The origin of the arc is unclear. \cite{higuchi_2014} speculate that the arc may have been generated following a collision between two molecular clouds based on the arc's morphological similarity to the structure generated in numerical simulations of cloud-cloud collisions \citep[e.g.][]{Habe1992, Takahira2014, haworth_2015}. An alternative hypothesis however, is that the arc is generated by stellar feedback. If confirmed, this could indicate that \brick \ is perhaps more active in its star formation than previously thought. In this work, we build on the analysis of \citet{henshaw_2019}, and introduce new observations from the Karl Jansky Very Large Array (VLA),\footnote{The VLA radio telescope is operated by the National Radio Astronomy Observatory (NRAO). The National Radio Astronomy Observatory is a facility of the National Science Foundation operated under cooperative agreement by Associated Universities, Inc.} to help test this hypothesis, finding that the morphology, dynamics, and energetics of the arc are all consistent to within a factor of a few of those predicted for a simple analytical model of an expanding bubble driven by the wind from a high-mass star. The paper is organised as follows. In Section~\ref{data} we describe the data used in this work, both from \citet{henshaw_2019} and our VLA observations. In Section~\ref{results} we outline our main results. Finally in Sections~\ref{discussion} and \ref{conclusions} we discuss our findings and outline our conclusions, respectively.

\section{Data}\label{data}

\subsection{ALMA data and {\sc ScousePy} decomposition}\label{data:alma}

\begin{figure*}
\begin{center}
\includegraphics[trim = 50mm 15mm 55mm 30mm, clip, width = 0.98\textwidth]{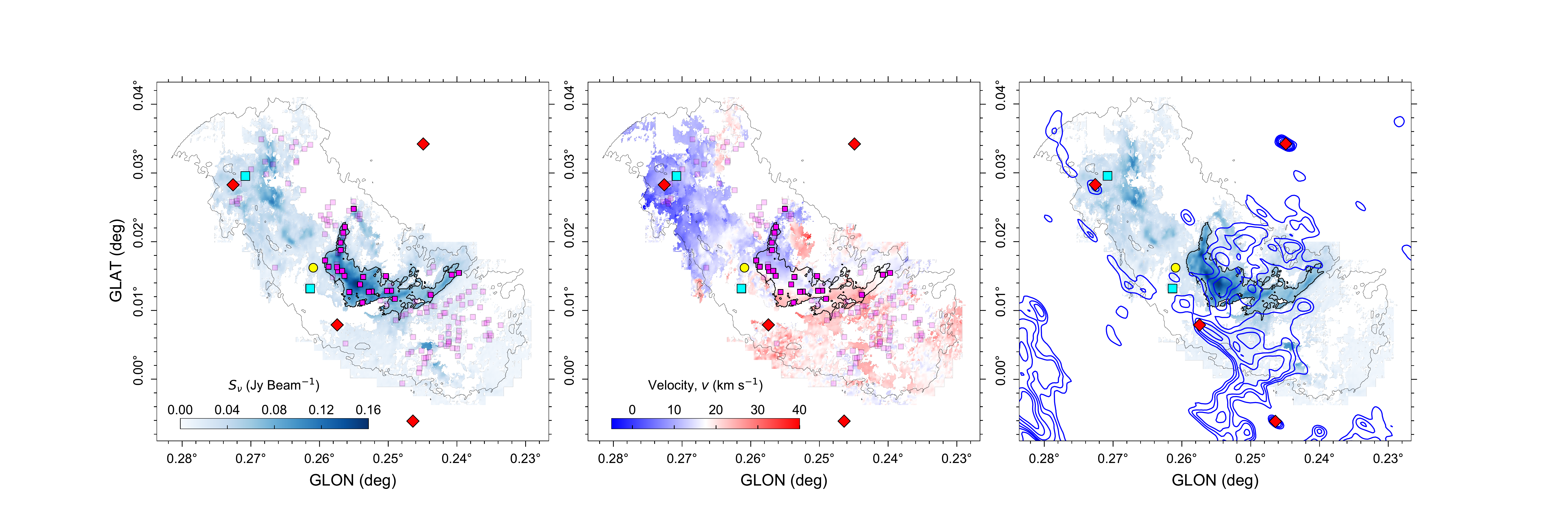}
\end{center}
\caption{Left: The peak flux distribution associated with the arc's parent sub-cloud identified in Paper~{\sc i}. The colour scale shows the peak amplitude of all Gaussian components associated with the arc's parent sub-cloud (derived from the fitting of the HNCO data). The thick black contour highlights the arc itself. The thin black contour shows the boundary of \brick \ estimated from the integrated emission of HNCO. We overlay the location of the H$_{2}$O maser identified by \citet{lis_1994} as a yellow circle and the additional H$_{2}$O masers identified by \citet{Lu2019} as cyan squares. H\,{\sc ii} region candidates from \citet{rodriguez_2013} are shown as red diamonds. Purple squares indicate the locations of class\,{\sc i} CH$_{3}$OH masers and maser candidates identified by \citet{mills_2015}. Transparent squares are those which lie outside of a $\pm6$\,\kms \ velocity range around a 2-D velocity plane fitted to the {\sc acorns} data (see text). Centre: The corresponding centroid velocity map of the arc's parent sub-cloud. The symbols are equivalent to those in the left panel. Right: The peak flux distribution with the VLA radio continuum data overlaid as blue contours. Contours start at $3\sigma$ ($\sigma=0.15$\,mJy\,beam$^{-1}$), then 5, 7, 10, 15, and 20$\sigma$ (Butterfield et al. in preparation).}
\label{Figure:arc_LS}
\end{figure*}

This paper makes use of the ALMA Early Science Cycle 0 Band 3 observations of G0.253+0.016 originally presented in \citet{rathborne_2014, rathborne_2015}. We summarise the observations here, but refer the reader to the aforementioned papers for a more extensive description. The ALMA 12\,m observations cover the full $3^{\prime}\times1^{\prime}$ extent of the cloud using a 13-point mosaic. Here, we use emission from the $4(0,4)-3(0,3)$ transition of HNCO, which has proved fruitful to study the internal structure and dynamics of the cloud \citep{rathborne_2015, federrath_2016, henshaw_2019}. \citet{rathborne_2015} combine these data with single dish observations from the Millimetre Astronomy Legacy Team 90 GHz Survey (MALT90; \citealp{foster_2011, jackson_2013}) obtained with the Mopra 22 m telescope to recover the extended emission filtered out by the interferometer. The spatial and spectral resolution are 1.7 arcsec and 3.4\,\kms, respectively. Throughout this paper we adopt a distance to the Galactic Centre of $8.178\pm0.013$\,kpc; \citealt{gravity_2019} and assume that \brick \ is located at this distance \citep{Nogueras-Lara2021a}. The corresponding physical resolution of these data is therefore $\approx 0.07$\,pc. The rms noise per 3.4\,\kms \ resolution element is 0.8\,mJy\,beam$^{-1}$. 

\citet{henshaw_2019} further process these data with the {\sc ScousePy} and {\sc acorns} algorithms \citep[Agglomerative Clustering for ORganising Nested Structures;][respectively]{henshaw_2016, henshaw_2019}, and again we summarise the procedure here, referring readers to the original paper for details. First, we use {\sc ScousePy} to decompose the spectral line emission into a set of discrete Gaussian components; we fit a total of $\sim450000$ Gaussian components to $\sim130000$ spectra (see Figure\,2 of \citealp{henshaw_2019}). We next use {\sc acorns} to cluster the Gaussian emission features identified by {\sc ScousePy} into hierarchical velocity-coherent regions. Out of the forest of clusters that {\sc acorns} identifies, four of them dominate the emission profile of G0.253+0.016 (as it appears in HNCO emission), accounting for $>50$ per cent of the detected Gaussian components. Of these four clusters, or trees as they are referred to in \citet{henshaw_2019} (owing to the dendrogram nomenclature), two account for the overall physical appearance of G0.253+0.016. The emission associated with the first, the ``main'' component, is qualitatively most similar in appearance to G0.253+0.016 as it appears in dust continuum emission \citep[][see their sect. 4.2]{henshaw_2019}. The emission profile of the second component is clearly associated with the arc focused on here, which previously had been detected in other works in different molecular species \citep{higuchi_2014, mills_2015}. This finding therefore served as the first evidence that the arc was coherent both in (projected) space and in velocity. In this work, we make use of the data products output from {\sc ScousePy} and {\sc acorns} related to this latter cloud component to investigate the origins of the arc. In the remainder of the paper, we refer to the component identified by {\sc acorns} as the parent sub-cloud of the arc.

\subsection{VLA data}

The VLA observations presented in this paper were taken in C Band (4--8 GHz) with the C array configuration (5\arcsec~resolution). The observations were taken in four separate observing runs, in June 2017, with a cadence of $\sim$2 days between observations. The observations targeted 6 separate fields, 2 hours on source per field. The observations used J1331+3030 (3C286) as the bandpass calibrator and J1820-2528 as the phase calibrator. The phase calibrator was observed every 35 minutes during the observations. The observations were also set up to observe the full stokes parameters and therefore we used J1407+2827 as the polarization leakage calibrator.  The observations were processed using the Common Astronomy Software Application (CASA)\footnote{\href{url}{http://casa.nrao.edu/}} pipeline, provided by NRAO, to calibrate the data. The continuum data combines the 4--8 GHz frequency coverage (3.8 GHz total bandwidth) of the C band observations. The continuum data used all 4 observing runs which were combined in the imaging stage of the data reduction. The observations were cleaned using the CASA task \texttt{tclean}. The image was cleaned non-interactively down to a threshold of 0.01 mJy. We used Briggs weighting of 0.5 to improve the sensitivity and resolution of the image. The data was cleaned using the 'multi-scale, multi-frequency synthesis' (deconvolver=`mtmfs', specmode=`mfs') with scales of 0, 4, and 16 pixels to account for the large scale structures present in the field. The synthesised beam size is $6\farc4\times2\farc9$ with a position angle $-2\fdeg5$. The rms noise (estimated from emission-free regions) is $0.15$\,mJy\,beam$^{-1}$. 

The radio recombination line data presented in this paper combined the H114$\alpha$, H113$\alpha$, H110$\alpha$, H109$\alpha$, H101$\alpha$, H100$\alpha$, and H99$\alpha$ transitions. The radio continuum was subtracted in the uv-plane, using the CASA task \texttt{uvcontsub}, before any imaging was done.
Each radio recombination transition was cleaned individually using the CASA task \texttt{tclean} by combining the four observing runs during the imaging process. All recombination line transitions were imaged using the same \texttt{tclean} parameters: 1 km s$^{-1}$ spectral resolution, 6$''$ $\times$ 12$''$ restoring beam size, velocity range of -40 to 99 km s$^{-1}$. The images were cleaned non-interactively using a set noise threshold level of 1 mJy and natural weighting to obtain the best sensitivity possible.
The cleaned images were then averaged together using the CASA task \texttt{immath} to improve the signal to noise in the image. 

\section{Results}\label{results}

\subsection{Morphology and kinematics}\label{results:arc}

We present a map of the arc in the left-hand panel of Figure~\ref{Figure:arc_LS}. The colour scale in this image refers to the peak amplitude of emission features extracted using {\sc ScousePy} (\S~\ref{data:alma}) from the HNCO data \citep{henshaw_2019}. The arc can be clearly identified in this map as the ridge of emission towards the centre of the cloud (highlighted by the thick black contour). 

We highlight several features of interest in the map. First, the yellow circle denotes the position of the H$_{2}$O maser identified by \citet[see also \citealp{Lu2019b}]{lis_1994}, which remains the only confirmed site of embedded star formation within \brick \ \citep[see also;][]{Walker2021}. The red diamonds are the locations of H\,{\sc ii} regions and H\,{\sc ii} region candidates in close projected proximity to \brick \ \citep[][though note that \citealp{mills_2015} argue that the sources within the cloud are spatially filtered peaks of more extended emission, as is also seen in the 5 GHz data presented here]{rodriguez_2013}. 

\citet{mills_2015} found a number of class~{\sc i} CH$_{3}$OH masers and maser candidates located throughout \brick. Rather than tracing the locations of on-going star formation, these most likely trace regions of shocked gas emission \citep{mills_2015}. To investigate whether any maser sources are associated with the arc, we can compare the positions and velocities of the masers with those of the arc. To do this, we first fit the velocity field of the arc parent cluster (see Figure~\ref{Figure:arc_LS}) with a bivariate polynomial (cf. \citealp{federrath_2016, henshaw_2019}). The velocity field displayed in Figure~\ref{Figure:arc_LS} shows a clear gradient, which increases from $\sim0$\,\kms \ in the (Galactic) north-east to $\sim25$\,\kms \ in the south-west of the the cloud, which we fit using 
\begin{equation}
    v_{\rm mod} = v_0 + \mathcal{G}_{l} l + \mathcal{G}_{b} b
\end{equation}
where $v_{0}$ is the systemic velocity of the source, $l$ and $b$ are the Galactic longitude and latitude, and $\mathcal{G}_{l}$ and $\mathcal{G}_{b}$ are the longitudinal and latitudinal components of the velocity gradient, respectively. The best-fit parameters are $v_{0}=14.7$\,\kms, and (converting from degrees to physical units) $\mathcal{G}_{l}=1.2$\,\kms\,pc$^{-1}$, and $\mathcal{G}_{b}=-1.0$\,\kms\,pc$^{-1}$. We then cross reference the maser catalogue of \citet{mills_2015} against this function, identifying all masers that lie in the range $v_{\rm mod} \pm 6$\,\kms. This velocity limit represents $\approx2$ resolution elements in the ALMA HNCO data. We highlight the 24 masers that are associated with the arc as opaque magenta squares in Figure~\ref{Figure:arc_LS} (masers outside of this velocity range are shown as semi-transparent magenta squares). These masers clearly follow the curvature of the arc, highlighting the association between the arc and the shocks traced by the class~{\sc i} CH$_{3}$OH masers. In addition to these masers, \citet{mills_2015} noted the presence of more extended, non-masing CH$_{3}$OH emission toward the arc. This is suggested to be quasithermal or 'quenched' emission \citep{menten_1991,mehringer_1997}, indicative of higher gas densities in this region. 

\begin{figure*}
\begin{center}
\includegraphics[trim = 10mm 10mm 10mm 10mm, clip, width = 0.85\textwidth]{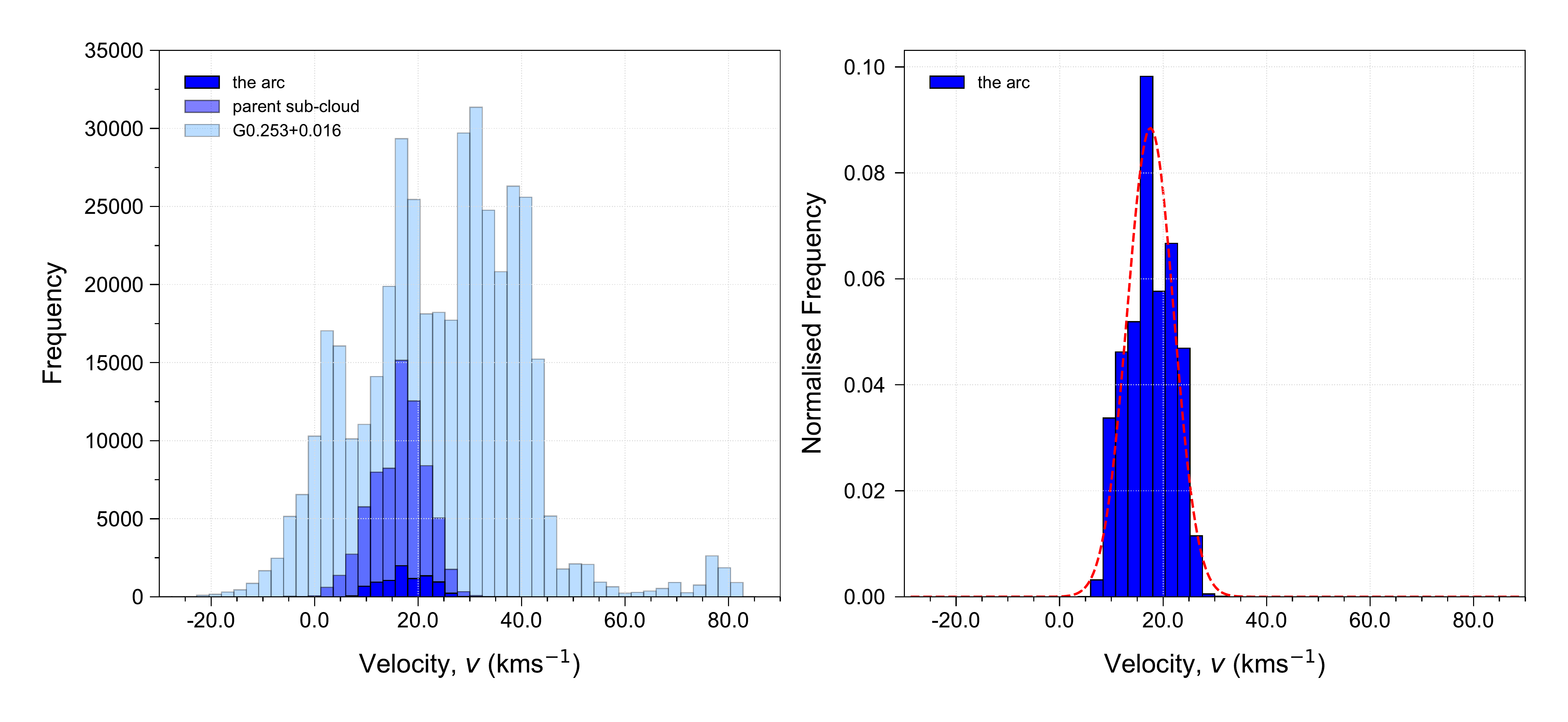}
\end{center}
\caption{Left: Histogram of the centroid velocity measurements associated with \brick. The light blue histogram displays all velocities extracted from the HNCO data across \brick, medium blue refers to the velocities of the arc's parent sub-cloud, and dark blue is a histogram of the arc velocities. Right: Normalised histogram of the centroid velocities associated with the arc. We overlay a Gaussian fit to the histogram (red dashed line). The mean velocity is $\langle v \rangle = 17.6\,\pm\,4.5$\,\kms, where the uncertainty here refers to the standard deviation of the distribution. }
\label{Figure:arc_vel_histo}
\end{figure*}

In the right-hand panel of Figure~\ref{Figure:arc_LS}, we present the 5\,GHz radio continuum emission observed with the VLA (blue contours). A striking feature of this emission is that it appears to fill the cavity traced by the arc. The emission within the arc cavity also connects in projection to a ridge of radio continuum emission that traces the outer (Galactic) eastern edge of the cloud. This latter ridge has been noted in earlier studies and has been attributed to the ionising influence of a known O4-6 supergiant located towards the (Galactic) south-east of the cloud \citep{mauerhan_2010, mills_2015}.

Figure~\ref{Figure:arc_vel_histo} is a histogram of the centroid velocity information extracted in \cite{henshaw_2019}. The left panel shows the distribution of centroid velocities for three distinct components. The dark blue histogram shows the arc itself, defined as the  region enclosed by the thick black contour in Figure~\ref{Figure:arc_LS}. For comparison, the medium blue histogram shows the arc's parent sub-cloud, and the light blue histogram shows all of \brick~\citep{henshaw_2019}. A Gaussian fit to the dark blue histogram (red dashed Gaussian in Figure~\ref{Figure:arc_vel_histo}) gives a mean velocity of $\langle v \rangle = 17.6$\,\kms with a standard deviation of $4.5$\,\kms.

\subsection{A simple geometrical model}\label{results:toy}

To better understand the morphology and dynamics of the arc we construct a simple model of a tilted ring projected on the plane of the sky \citep[cf.][]{Lopez-Calderon2016, Callanan2021}. The model is described by five free-parameters: i \& ii) the coordinates of the ring centre on the plane of the sky, $\{l_{0}, b_{0}\}$, iii) the radius of the ring, $R_{\rm arc}$; and iv \& v) two angles, $\beta$, $\gamma$, that describe the orientation of the ring relative to the plane of the sky \citep[inclination and position angle, see][]{Callanan2021}. Formally, we describe the shape of the ring by constructing a local Cartesian coordinate system centred on the ring, with $\hat{x}$ along the line of sight, and $\hat{y}$ and $\hat{z}$ aligned with Galactic longitude and latitude. We begin with a ring lying in the $xy$ plane of this coordinate system (i.e.\ edge-on from our point of view, and at constant Galactic latitude), whose coordinates can be expressed parametrically as $\mathbf{r} = (R_{\rm arc} \cos\theta, R_{\rm arc} \sin \theta, 0)$ with $\theta \in [0,2\pi)$. The angles $\beta$ and $\gamma$ then represent rotations about the $y$ and $x$ axes of this coordinate system\footnote{We need not consider rotations about the $z$ axis for reasons of symmetry.}, so the coordinates of the ring become $\mathsf{R}_y(\beta) \mathsf{R}_x(\gamma) \mathbf{r}$, where $\mathsf{R}_x$ and $\mathsf{R}_y$ are the usual rotation matrices for rotations about the $x$ and $y$ axes:
\begin{eqnarray}
\mathsf{R}_y(\beta)& = &
\begin{bmatrix}
\cos \beta & 0 & \sin \beta\\
0 & 1 & 0 \\
-\sin\beta  & 0 & \cos\beta
\end{bmatrix}
\\
\mathsf{R}_x(\gamma) &= &
\begin{bmatrix}
1 & 0 & 0 \\
0 & \cos \gamma & -\sin\gamma \\
0 & \sin\gamma  & \cos\gamma
\end{bmatrix}
.
\end{eqnarray}

To find the parameters that best describe the arc, we minimise the distance between the image pixels that we identify as being in the arc and the projected arc model. Formally, our procedure is as follows. For any proposed vector of parameters $\mathbf{P}$ describing the arc, we first compute the projected position of the arc in the Cartesian coordinate system defined by the observed image; we denote this projected position $\left(x_{\mathbf{P}}(\theta), y_{\mathbf{P}}(\theta)\right)$, where $\theta$ is a parametric variable that varies from 0 to $2\pi$. The data to which we fit this model consists of the set of $N$ pixels in the image that we have identified as being part of the arc; let $(x,y)_i$ for $i=1\ldots N$ denote the positions of the centres of these pixels in the image coordinate system. For each pixel $i$ we define the distance to any point on the model arc by
\begin{equation}
d_{i,\mathbf{P}}(\theta) = \sqrt{ \left[x_{\mathbf{P}}(\theta) - x_i\right]^2 + \left[y_{\mathbf{P}}(\theta) - y_i\right]^2 } ,
\end{equation}
and we further define $d_{\mathrm{min},i,\mathbf{P}}$ as the minimum of $d_{i,\mathbf{P}}(\theta)$ on the domain $\theta = [0,2\pi]$, i.e.\ $d_{\mathrm{min},i,\mathbf{P}}$ is the minimum distance from the centre of pixel $i$ to any point on the arc. We define our goodness of fit statistic for a proposed set of model parameters $\mathbf{P}$ by $\chi^2(\mathbf{P}) = \sum_{i=1}^N d_{\mathrm{min},i,\mathbf{P}}$, i.e.\ the goodness of fit of the model is simply the sum of the squared minimum distances between the arc pixels in the image and the projected arc produced by a given set of model parameters. We find the set of parameters $\mathbf{P}$ that minimise this objective function using a standard Levenberg-Marquardt minimisation method \citep{Newville2014}. 

\begin{figure*}
\begin{center}
\includegraphics[trim = 25mm 5mm 40mm 25mm, clip, width = 0.85\textwidth]{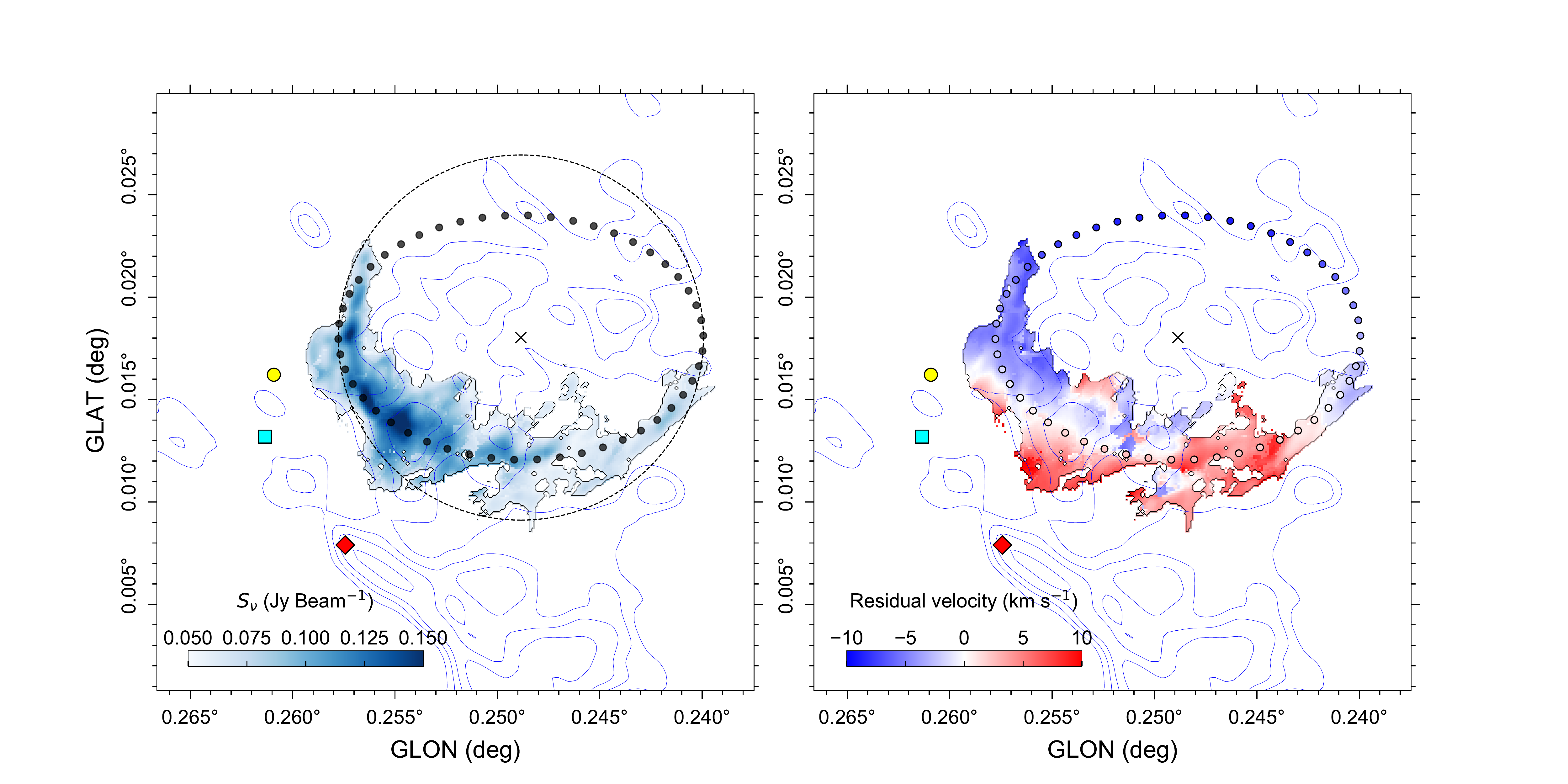}
\end{center}
\caption{Left: The colour-scale indicates the peak amplitude of the Gaussian components associated with the arc (derived from the fitting of the HNCO data). The thick black dotted circle indicates our best-fitting toy model of an expanding ring. It is centred on $\{l_{0},b_{0}\}=\{0.248\degr,\,0.018\degr\}$ and has a radius $R_{\rm arc}=32$\arcsec \ or $R_{\rm arc}=1.3$\,pc. The dashed circle has an equivalent radius and is shown for reference. Right: The velocity field of the arc after subtracting the bulk motion of the arc's parent sub-cloud. The dotted circle once again shows the geometry of our best-fitting toy model, however here the colour of the dots indicates the expansion of the ring (see text for details). The contours are equivalent to those in Figure~\ref{Figure:arc_LS}.}
\label{Figure:arc_SS}
\end{figure*}

Our best-fitting model geometry is displayed in Figure~\ref{Figure:arc_SS}, where it is overlaid on maps of the peak amplitude and gradient-subtracted velocity field (see \S~\ref{results:arc}) of the arc. The circular model forms an ellipse when projected on the plane of the sky. It is centred on $\{l_{0},b_{0}\}=\{0\fdg248,\,0\fdg018\}$ and has a radius $R_{\rm arc}=32$\arcsec \ or $R_{\rm arc}=1.3$\,pc.\footnote{If we would have simply fitted the arc as a circle on the plane of the sky, we would have obtained $\{l,b\}=\{0\fdg250,\,0\fdg018\}$ and a radius of $R_{\rm arc}=25.8$\arcsec \ or $R_{\rm arc}=1.0$\,pc.}

\begin{figure}
\begin{center}
\includegraphics[trim = 10mm 33mm 10mm 50mm, clip, width = 0.48\textwidth]{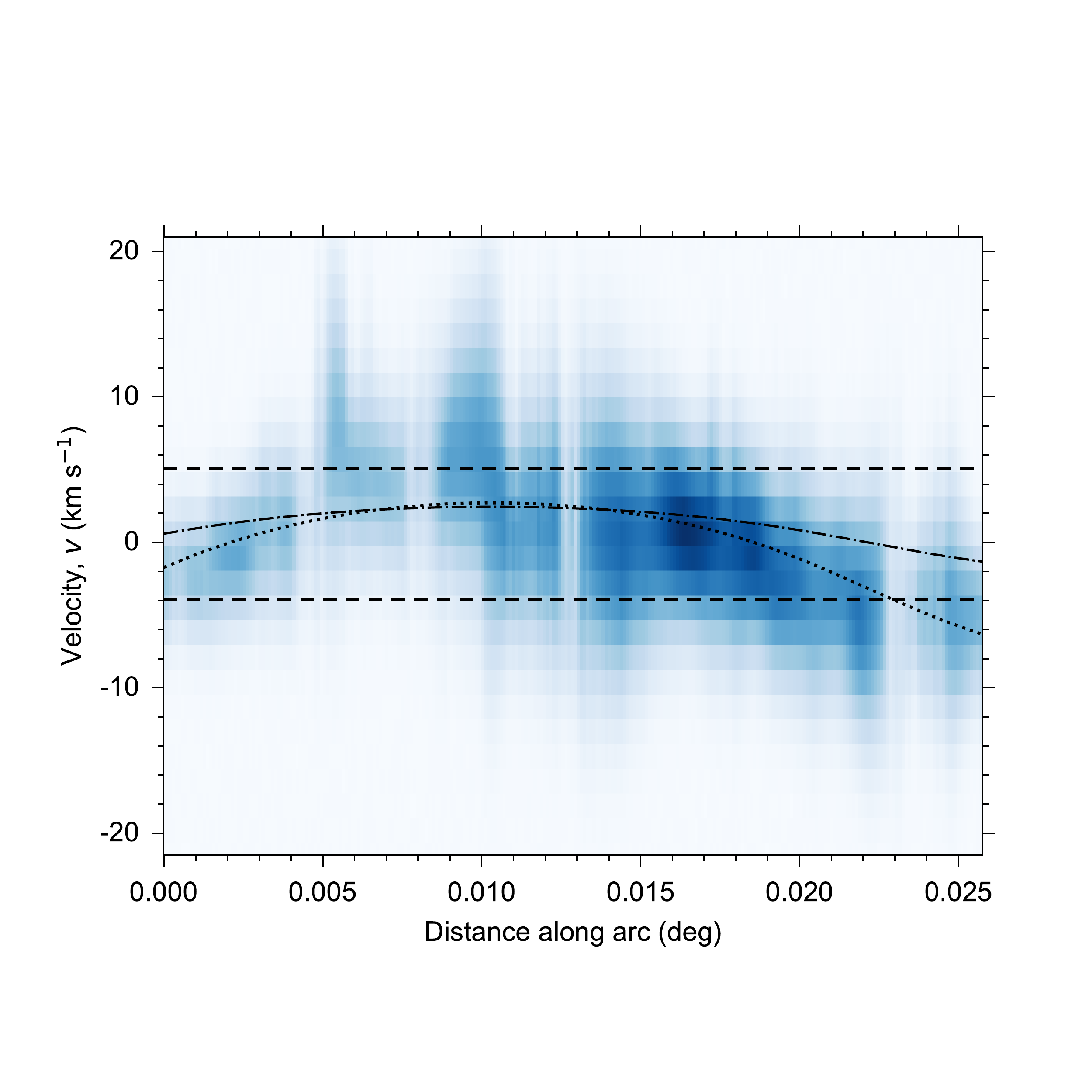}
\end{center}
\caption{A position-velocity diagram extracted along the dotted ellipse presented in Figure~\ref{Figure:arc_SS}. The 0.0 location is taken to be the lowest Galactic longitude point on the arc. The colour scale reflects the peak amplitude of the HNCO emission. The lines represent different models for the kinematics of the arc velocity field presented in the right panel of Figure~\ref{Figure:arc_SS}. The horizontal dashed lines represent the most simplistic approach to estimating the expansion velocity, and reflect the half-width at half-maximum of the gradient-subtracted velocity distribution (see text for details), $v_{\rm exp}=4.2$\,\kms. The dot-dashed and dotted lines correspond to the model velocity fields described in \S~\ref{results:toy}. The former of these models has a constant expansion velocity of $v_{\rm exp}=3.3$\,\kms. The latter also has constant expansion velocity, this time $v_{exp}=7.9$\,\kms, but the model also includes a constant line-of-sight velocity of $v_{\rm 0,arc}=-3.1$\,\kms. }
\label{Figure:pvarc}
\end{figure}

The gradient-subtracted velocity field (see \S~\ref{results:arc}) presented in the right-hand panel of Figure~\ref{Figure:arc_SS} is quite complex. Broadly speaking, the velocities transition from blue- to red- and back to blue-shifted emission again in the azimuthal direction. Gradients in the radial direction further complicate this picture. However, the azimuthal trend may be produced by the expansion of the arc. We can verify this with our toy model. First, we assume that the arc is expanding radially and second, that the expansion velocity is constant in azimuth in the plane of the arc. Having fixed the geometry, we perform another least squares fit to determine the expansion velocity, $v_{\rm exp}$, that best describes the velocity field of the arc. We do this in two ways. In the first method, we include only the expansion velocity as a free-parameter in the model. In the second method, we introduce a constant in addition to the expansion velocity that represents the systemic line-of-sight velocity of the arc, $v_{\rm arc, 0}$. For the former we derive $v_{\rm exp}=3.3$\,\kms. For the latter, we derive $v_{\rm exp}=7.9$\,\kms and $v_{\rm arc, 0}=-3.1$\,\kms. The introduction of the additional free-parameter in the second method leads to the factor of $\sim2$ change in the modelled expansion velocity. This latter model is displayed as the coloured dots in the right-hand panel of Figure~\ref{Figure:arc_SS} (the colour scale of the dots matches that of the background velocity field). Finally, we introduce a ``control'' estimate of the expansion velocity by simply fitting a Gaussian to the distribution of gradient-subtracted centroid velocities shown in Figure~\ref{Figure:arc_LS}. We then estimate the expansion velocity as the half-width at half-maximum (HWHM) of this distribution, finding $v_{\rm exp}=4.2$\,\kms. Each of these estimates is highlighted in Figure~\ref{Figure:pvarc}, which is a position-velocity diagram extracted along the (partial) ellipse shown in Figure~\ref{Figure:arc_SS} (the 0.0 location is taken to be the lowest Galactic longitude point on the arc). The dot-dashed line reflects our kinematic model with $v_{\rm exp}=3.3$\,\kms, the dotted line represents the model with $v_{exp}=7.9$\,\kms, and the horizontal lines represent the HWHM approach with $v_{\rm exp}=4.2$\,\kms. 

The uncertainties in this modelling approach are considerable, and the velocity field of the arc is more complicated than that produced by this simplified model. Nonetheless, this simple approach demonstrates the plausibility that the morphology of the arc, as well as its dynamics, may be interpreted as an expanding shell. For the sections that follow, we propagate the uncertainties associated with this modelling into our calculations. We use the mean of the expansion velocities as our fiducial estimate, but retain the upper and lower limits for further calculations, $v_{\rm exp}=5.2^{+2.7}_{-1.9}$\,\kms. Under these assumptions, we can estimate the dynamical age of the arc,
\begin{equation}
t_{\rm dyn} = \frac{R_{\rm arc}}{v_{\rm exp}} \,.
\end{equation}
With our best-fitting values $R_{\rm arc}=1.3$\,pc and $v_{\rm exp}=5.2^{+2.7}_{-1.9}$\,\kms \ the estimated dynamical age is $t_{\rm dyn}\approx2.4^{+0.8}_{-1.4}\times10^{5}$\,yr (assuming a constant expansion velocity). 

\subsection{Mass, energy, and momentum}\label{results:energetics}

With an estimate of the expansion velocity we can estimate the energy and momentum associated with the arc. To do this we first estimate a mass using dust continuum emission. We derive the total mass of the arc within the black contour presented in Figure~\ref{Figure:arc_SS} from the 3\,mm dust continuum emission from ALMA Cycle 0, first presented by \citet{rathborne_2014}:
\begin{equation}
M_{\rm arc}=\frac{d^{2}S_{\nu}R_{\rm g2d}}{\kappa_{\nu}B_{\nu}(T_{\rm d})}, 
\label{Eq:mass_calc}
\end{equation}
where $d$ is the distance to the source, $S_{\nu}$ is the integrated flux density (in Jy), $R_{\rm g2d}$ is the gas-to-dust ratio, $\kappa_{\nu}$ is the dust opacity per unit mass at a frequency $\nu$, and $B_{\nu}(T_{\rm d})$ is the Planck function at a dust temperature, $T_{\rm d}$. We adopt a dust opacity per unit mass $\kappa_{\nu}=~\kappa_{0}(\nu/\nu_0)^{\beta}$ with $\kappa_{0}=0.899$\,cm$^{2}$g$^{-1}$, valid for the moderately coagulated thin ice mantle dust model of \citet{ossenkopf_1994} with densities of $10^{6}$\,cm$^{-3}$ at $\nu_{0}=230$\,GHz. We adopt $\beta=1.75$ following \citet{battersby_2011}, giving an opacity $\kappa_{\nu}\approx0.21$\,cm$^{2}$g$^{-1}$ at a frequency of $\sim93$\,GHz. 

Two considerable sources of uncertainty in our mass estimate are the dust temperature and the gas-to-dust ratio, $R_{\rm g2d}$. For the former, \brick \ overall shows low dust temperatures of the order $\sim20$\,K \citep{longmore_2012, Tang2021}. \citet{marsh_2016} find that the dust associated with the arc consists of a cool ($<20$\,K) and a warm component (up to $\sim50$\,K). In terms of the gas temperature, \citet[][see also \citealp{ginsburg_2016, krieger_2017}]{mills_2018} also find evidence from HC$_{3}$N emission in \brick\ for two distinct components, one low-excitation, low-density ($n\sim10^3$\,cm$^{-3}$; $T\sim25-50$\,K) and one high-excitation, high-density ($n\sim10^5$\,cm$^{-3}$; $T\sim60-100$\,K). The gas temperature in Galactic Centre clouds is typically higher than the dust temperature \citep{krieger_2017} and modelling indicates that even at densities of $10^{5}$\,cm$^{-3}$, the gas and dust are unlikely to be in thermal equilibrium \citep{clark_2013}. The uncertainty on the dust temperature is most likely a factor of 2. Moreover, given that the metallicity in the Galactic Centre is approximately twice solar \citep{Mezger1979,Feldmeier-Krause2017, Schultheis2019, Schultheis2021}, the gas-to-dust ratio is likely lower by a similar factor \citep{longmore_2013, Giannetti2017}.

Combining the above uncertainties, we estimate that the arc has a mass of $M_{arc}\sim2700^{+3000}_{-1400}$\,M$_{\odot}$, where the fiducial value corresponds to $T=50$\,K and $R_{\rm g2d}=100$ (or $T=25$\,K and $R_{\rm g2d}=50$). We caution that this still likely represents a strict upper limit to the mass of the arc because there are multiple velocity components along the line-of-sight in this location, which are not accounted for in mass derivations from continuum observations. Importantly, the arc spatially overlaps with the dominant sub-cloud in \brick, which likely contains most of the mass \citep{henshaw_2019}. Therefore, although the uncertainty on the mass derived from continuum observations is of the order a factor of $\sim2$, this additional consideration means that the uncertainty could be higher. 

With estimates for the mass and expansion velocity in hand we can now estimate the kinetic energy and momentum of the arc using
\begin{equation}
E_{\rm arc}=\frac{1}{2}M_{\rm arc}v_{\rm exp}^{2}
\end{equation}
and 
\begin{equation}
p_{\rm arc}=M_{\rm arc}v_{\rm exp},
\end{equation}
finding $E_{\rm arc}\sim0.7^{+2.8}_{-0.6}\times10^{48}$\,erg and $p_{\rm arc}\sim1.4^{+3.1}_{-1.0}\times10^{4}$\,M$_{\odot}$\kms, respectively. We discuss these values in more detail in \S~\ref{discussion}.

\subsection{On the nature of the radio emission and the association between the arc and the ionised gas}\label{results:hii}

\begin{figure}
\begin{center}
\includegraphics[trim = 5mm 10mm 0mm 0mm, clip, width = 0.48\textwidth]{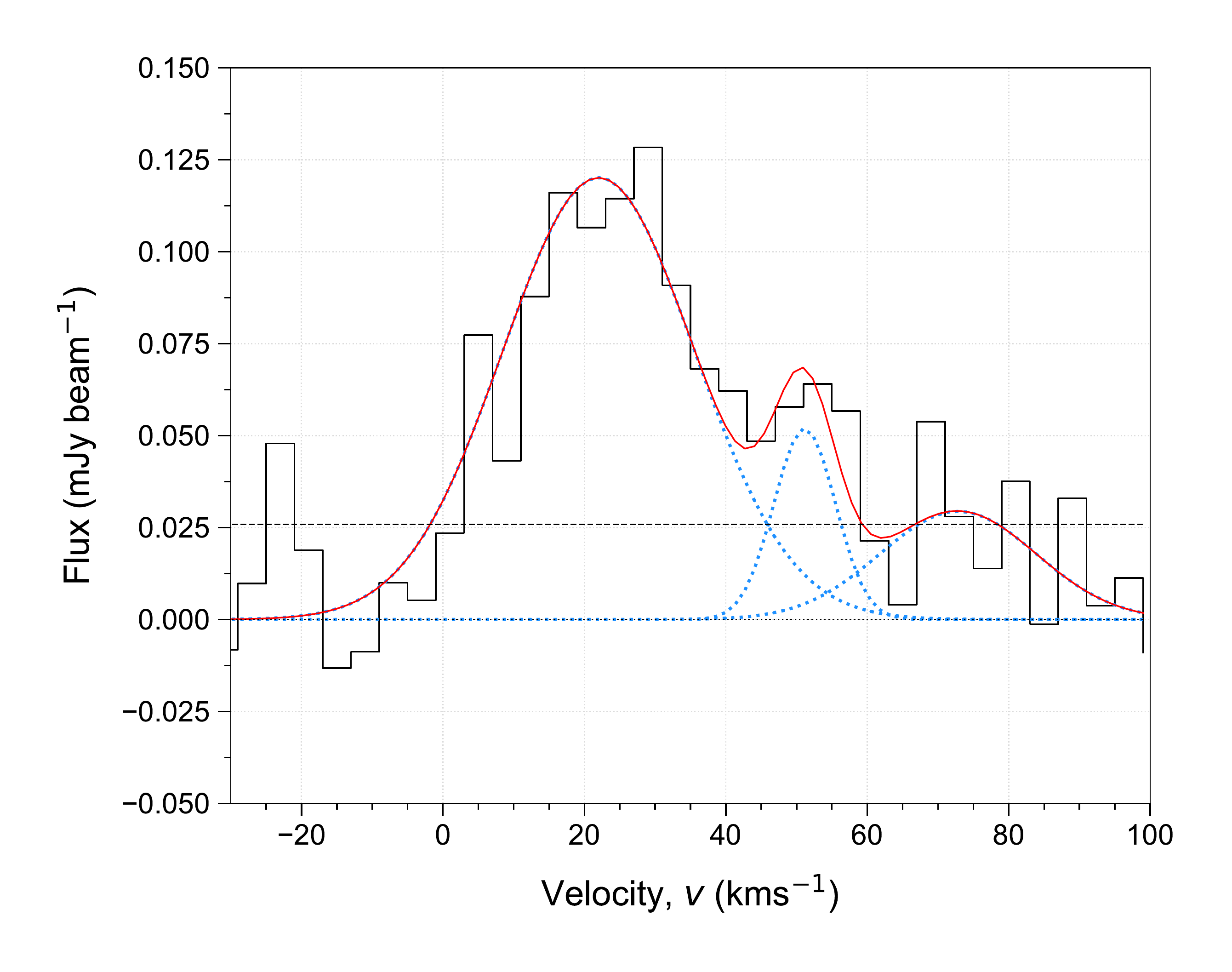}
\end{center}
\caption{Radio recombination line spectrum extracted from within the circle presented in Figure~\ref{Figure:arc_SS}. This spectrum was created by stacking a total of seven radio recombination lines, namely H114$\alpha$, H113$\alpha$, H110$\alpha$, H109$\alpha$, H101$\alpha$, H100$\alpha$, and H99$\alpha$. The horizontal dotted line indicates the 0.0 line and the horizontal dashed line indicates $3.0\times\sigma_{\rm rms}$ ($0.026$\,mJy\,beam$^{-1}$). The red curve indicates a three component Gaussian fit to the data. The component at lower velocities has a centroid velocity of $22.0\pm1.4$, closely matching the velocity of the molecular component of arc (Figure~\ref{Figure:arc_vel_histo}). }
\label{Figure:RRL}
\end{figure}

Radio continuum emission is detected throughout the arc cavity in projection (Figure~\ref{Figure:arc_LS}). However, as discussed in \S~\ref{results:arc}, the emission extends further to the (Galactic) south and east. While it is certainly possible that the radio continuum emission is physically related to the arc, projection effects may be important. To investigate whether the ionised gas is physically associated with the molecular arc, we extract a radio recombination line (RRL) spectrum from the region marked with a dotted circle in Figure~\ref{Figure:arc_SS}. In practice, we stack the emission from a total of seven RRL transitions, namely, H114$\alpha$, H113$\alpha$, H110$\alpha$, H109$\alpha$, H101$\alpha$, H100$\alpha$, and H99$\alpha$. The resulting spectrum is displayed in Figure~\ref{Figure:RRL}. In addition to stacking, we have smoothed the native spectral resolution of the stacked spectrum by a factor of 4 to further increase the signal-to-noise. We fit the smoothed spectrum using a multi-component Gaussian model using the stand-alone fitter functionality of {\sc ScousePy} \citep{henshaw_2019}. This procedure uses derivative spectroscopy to determine the number of emission features within each spectrum and their properties \citep[i.e.\ their peak amplitude, velocity centroid, and width;][]{Lindner2015,Riener2019}. Using a Gaussian smoothing kernel of standard deviation 1.5 channels, and ensuring that all identified components are above a signal-to-noise ratio of 3, this method predicts a three component model. The brightest component has a centroid velocity of $22.0\pm1.4$\,\kms and has a velocity dispersion of $13.6\pm1.5$\,\kms. This velocity is redshifted with respect to the mean of the arc centroid velocity distribution ($17.6$\,\kms), but is consistent to within one standard deviation and is importantly inconsistent with the other sub-clouds associated with \brick \ \citep{henshaw_2019}. Note that the combination of the broad lines, spectral smoothing, and the narrow bandwidth make it difficult to determine if the two lower brightness emission features are significant. However, they are located at higher velocity and are therefore not relevant here. The consistency in velocity between the RRL emission and the molecular gas tracing the arc, in addition to the spatial relationship between the radio continuum emission and the arc cavity, leads us to conclude that the molecular gas and ionised gas are most likely related.

To help better understand the nature of the ionised gas we estimate the electron density, recombination time, and Lyman continuum ionising flux. The morphological and kinematic match between the radio emission presented here (continuum and RRL emission, respectively) and the arc (Figure~\ref{Figure:arc_SS}) gives us confidence that the two are physically related. However, we note that \brick \ lies close in projection to both of thermal and non-thermal radio sources, in particular the arched radio filaments that are oriented perpendicular to the Galactic plane \citep{Morris1989, Yusef-Zadeh1989}. \brick \ also overlaps in projection with the prominent supernova remnant G0.30+0.00 \citep{kassim_1996, larosa_2000}, and an additional candidate supernova remnant lies directly to the Galactic west of the arc \citep{ponti_2015}. The contribution of non-thermal emission to the radio continuum flux may therefore be non-negligible. We, therefore, estimate the electron density, recombination time, and Lyman continuum ionising flux in two ways i) assuming that the radio continuum flux is produced entirely by free-free emission, which provides our upper limit; ii) using the RRL emission to self-consistently predict what the expected free-free continuum flux would be.

The total integrated continuum flux within the arc cavity (see the circle in Figure~\ref{Figure:arc_SS} is $\sim 80$\,mJy. This provides our strict upper limit on the free-free emission. The measured RRL integrated intensity in Figure~\ref{Figure:RRL} is 5.2 mJy\,\kms \ (4.6\,K\,\kms). Assuming that the RRLs are optically thin and in LTE \citep[typical departure coefficients $\beta_n$ are very close to unity for H99-114$\alpha$;][]{Storey1995} we can use equation 14.29 of \citet[][5th ed.]{Wilson2009} to derive the line-to-continuum ratio of the RRLs, $T_L/T_C$
\begin{equation}
    \frac{T_L}{T_C}
    \frac{\delta v}{\kms}
    =
    \frac{6.985\times10^{3}}{a(\nu, T_e)}
    \left[\frac{\nu}{\mathrm{GHz}}\right]^{1.1}
    \left[\frac{T_e}{\mathrm{K}}\right]^{-1.15}
    (1+y_{He})^{-1}
\end{equation}
where $a(\nu, T_e)$ is the Gaunt factor, assumed to be unity, and ${y_{He} = N(He+)/N(H+)}$, the ratio of helium to hydrogen ions, is assumed to be 0.1. We determine $T_L/T_C\approx2.5$\,\kms for $T_e=5000$\,K (see above). From this ratio we determine that the expected continuum flux is $\approx1$\,mJy (cf. $\sim80$\,mJy derived from the continuum). This calculation indicates that the continuum likely suffers contamination from non-thermal emission, and the estimated continuum flux from the RRL emission provides a lower bound to the contribution from free-free emission.  

The electron density within the shell (assuming that the ionised gas fills the volume of the shell bounded by the arc) is \citep{mezger_1967, rubin_1968}
\begin{equation}
n_{e}=2.3\times10^{6}\bigg[\frac{S_{\nu}}{\mathrm{Jy}}\bigg]^{0.5}\bigg[\frac{\nu}{\mathrm{GHz}}\bigg]^{0.05}\bigg[\frac{T_e}{\mathrm{K}}\bigg]^{0.175}\bigg[\frac{d}{\mathrm{pc}}\bigg]^{-0.5}\bigg[\frac{\theta}{\mathrm{arcsec}}\bigg]^{-1.5}\,\mathrm{cm}^{-3} ,
\end{equation}
where $S_{\nu}$ is the integrated flux density at a frequency $\nu$ (5\,GHz), $T_{e}$ is the electron temperature (which we assume to be $T_{e}=5\times10^{3}$\,K, relevant for the electron temperature in Galactic Centre H{\sc ii} regions; \citealp{lang_1997, deharveng_2000, law_2009}), $d$ is the source distance, and $\theta=2R=64\arcsec$ refers to the angular size of the source. The recombination time is $t_{\rm rec}=1/(n_{e}\alpha_{B})$, where $\alpha_{B}$ is the hydrogen recombination coefficient, which we assume is $\alpha_{B}=4.5\times10^{-13}$\,cm$^{3}$s$^{-1}$ (valid for an assumed temperature of $5\times10^{3}$\,K; \citealp{draine_2011}). For the lower and lower bounds on the free-free emission, we derive a range in electron density of $n_{e}\approx10\mhyphen93$\,cm$^{-3}$. The corresponding range in recombination time is $t_{\rm rec}\approx760\mhyphen7000$\,yr. 

The Lyman continuum photon injection rate needed to balance recombinations is
\citep{mezger_1967, rubin_1968}:
\begin{equation}
N_{\rm LyC}=8.40\times10^{40}\bigg[\frac{S_{\nu}}{\mathrm{Jy}}\bigg]\bigg[\frac{\nu}{\mathrm{GHz}}\bigg]^{0.1}\bigg[\frac{T_e}{10^{4}\mathrm{K}}\bigg]^{-0.45}\bigg[\frac{d}{\mathrm{pc}}\bigg]^{2}\,\mathrm{s}^{-1}.
\end{equation}
Inserting numerical values, we derive a range for the Lyman continuum ionising flux of $N_{\rm LyC}\approx10^{46.0}\mhyphen10^{47.9}$\,photons\,s$^{-1}$. The Lyman continuum photon rate gives us some insight into the type of source that may be driving this emission. Assuming that the emission is produced by a single zero-age main sequence star, the bounds of our derived $N_{\rm LyC}$ values correspond to stars of spectral type B1-O8.5, with corresponding masses $12\mhyphen20$\,M$_{\odot}$ \citep{panagia_1973, Smith2002, Martins2005, Armentrout2017}. We conclude that the driving source of the continuum may be a high-mass star. In the following sections we discuss whether such a star is the likely driving source of the arc.

\section{Discussion}\label{discussion}

In the case of massive stellar clusters ($M>10^{3}$\,M$_{\odot}$), the energetic processes are dominated by three main forms of feedback: ionising radiation, stellar winds, and supernovae \citep{Krumholz2014}. Stellar feedback plays an integral role in shaping the ISM and regulating star formation at the centre of the Galaxy \citep{kruijssen_2014b, Krumholz2017, Armillotta2019, Barnes2020, Tress2020, Sormani2020}. Although the star formation rate is low in the CMZ \citep{longmore_2013}, the Galactic Centre star-forming regions (e.g.\ Sgr\,B2 and Sgr\,A) are among the most luminous in the Milky Way. The results presented in the previous section, specifically the morphology and dynamics of the molecular arc and its apparent physical association with the ionised gas emission, suggest that the arc may be the result of stellar feedback. This conclusion is at odds with previous works suggesting that the arc may have been generated during a cloud-cloud collision \citep{higuchi_2014}. This conclusion is also in tension with the generally accepted view that \brick \ is largely quiescent, with only a single known site of confirmed active star formation \citep{Walker2021}. In the following sections, we discuss the possible origins of the arc, assuming that it is generated by stellar feedback, before addressing the question of whether or not we would expect to detect its progenitor star towards \brick. 

\subsection{Is the arc a shell swept up by the wind of an interloper star?}\label{discussion:bowshock}

One hypothesis that would be consistent with the quiescent picture of \brick, is that the arc represents a shell swept up by the wind of an interloper star. High-mass stars possess powerful winds and the CMZ is unique in our Galaxy in that there is a rich population of `field' high-mass stars distributed throughout \citep{mauerhan_2010, dong_2011, Clark2021}. The origin of this population is unclear. In general, the lifetimes of molecular clouds in the CMZ are short ($\sim1$\,Myr; \citealp{henshaw_2016c,jeffreson_2018}). Clouds are destroyed by powerful stellar feedback \citep{Barnes2020} and their emergent stellar populations contribute to the field. Another possibility is that some of this population results from the tidal stripping of, or from stellar interactions within the CMZ's massive clusters the Arches and Quintuplet \citep{habibi_2014}. Irrespective of their origins, the impact that these high-mass field stars have on the surrounding interstellar medium is not well understood \citep[although see][]{Simpson2018, Simpson2021}. 

We can crudely estimate the likelihood that the star represents an interloper using simplistic assumptions based on the known properties of the CMZ. If we take the approximate present day star formation of the CMZ, $\sim0.1$\,\solar\,yr$^{-1}$ \citep[which has been more or less constant over the past several Myr;][]{longmore_2013, barnes_2017}, and make the assumption that the vast majority of this star formation is confined to a torus with major and minor radii of $\sim100$\,pc and $\sim10$\,pc, respectively \citep{molinari_2011, kruijssen_2015, henshaw_2016}, the expected volumetric star formation rate is of the order $\sim0.5$\,\solar\,Myr$^{-1}$\,pc$^{-3}$. 
First consider a scenario where the interloper is an O star with a lifetime $\approx 4$ Myr. Assuming that a single $16-20$\,\solar \ star is produced for every $\sim$500\,\solar \ cluster produced (assuming a standard \citealt{kroupa_2001} initial mass function; IMF), the density of $16-20$\,\solar \ stars is $\rho_* = 1/250$\,pc$^{-3}$, and the expected number within the volume of \brick, assuming a cross sectional area $A\sim17$\,pc$^{2}$ and a depth $L=4.7$\,pc \citep{federrath_2016}, is $\langle N \rangle = AL\rho_* \approx 0.3$. This is high enough that we must consider the possibility that an interloper might be responsible for the arc. In the alternative scenario where the interloper is a B star, the expected number is even larger, since B stars are both more common and live longer.

Numerical simulations show that the winds from runaway O and B stars can sweep up a dense shell as they pass through molecular clouds \citep{Mackey2015}. It is tempting to speculate that such a star may have been exiled from the Arches or Quintuplet \citep{portegies-zwart_2010}. This possibility has been discussed in relation to both Sgr\,B1 \citep{Simpson2018} and the Sgr\,A-H group of H\,{\sc ii} regions \citep{Hankins2019}. The Arches cluster in particular is located to the (Galactic) west of \brick \ and has a projected distance of just $\sim20$\,pc. The Arches is a young ($2-4$\,Myr; \citealp{najarro_2004, martins_2008}) massive ($4-6\times10^{4}$\,\solar; \citealp{clarkson_2012}) cluster containing a large number of high-mass stars \citep{hosek_2015}.

To explore this hypothesis further, we can examine the size of the arc in more detail. As the relative velocity between the runaway star and the ambient medium increases, the characteristic size of the swept-up shell driven by the star's wind decreases \citep{Mackey2015}. The scale of the bow shock produced, the stand-off distance, is defined as the point where the momentum flux of the stellar wind balances the momentum flux of the ambient medium, and is given by \citep{baranov_1971,green_2019},
\begin{equation}
    R_{\rm st}=\sqrt{\frac{\dot{M}v_{\infty}}{4\pi \rho_{0}(v_{*}^{2}+c_{s}^{2})}}
    \label{eq:bow}
\end{equation}
where $\dot{M}$ is the stellar wind mass loss rate, $v_{\infty}$ is the terminal wind velocity, $\rho_{0}$ is the density of the ambient medium, $v_{*}$ is the velocity of the star with respect to the ambient medium and $c_{s}$ corresponds to the sound speed, in this case in the molecular phase. This is because the bow shock is expected to trap the ionization front for the strong wind and dense interstellar medium derived above \citep{maclow_1991, arthur_2006}, in which case the bow shock expands into molecular gas. 

Using Equation~\ref{eq:bow}, we can ask the question: \emph{what size shell could be produced by the type of high-mass star needed to stimulate the ionised emission observed within the arc cavity?} To address this question we first estimate the mass loss rate and terminal wind velocity of the high-mass star. The limiting case, i.e. the star that is capable of producing a shell with the largest radius, is given by the upper end of our mass limit derived in \S~\ref{results:hii}. For O stars which span the range of spectral types consistent with our estimated Lyman continuum photon rate of $N_{\rm LyC}=10^{47.9}$\,photons\,s$^{-1}$ (O9.5, O9, O8.5), \citet[][see their Table 1]{Martins2005} provide stellar masses ($M/M_{\odot}=\{16.46, 18.03, 19.82\}$), luminosities ($\mathrm{log} L/L_{\odot}=\{4.62, 4.72, 4.82\}$), and effective temperatures ($T=\{30488, 31524, 32522\}$\,K). We can use this information to determine the mass loss rate using the metallicity-dependent relationship described in \citet{Vink2001}. We derive mass loss rates for two metallicities (consistent with our mass calculations in \S~\ref{results:energetics}), namely solar and twice solar, finding $\dot M(Z/\mathrm{Z_{\odot}}=1)=\{0.3,0.4,0.7\}\times10^{-7}$\,\solar\,yr$^{-1}$ and $\dot M(Z/\mathrm{Z_{\odot}}=2)=\{0.5,0.8,1.2\}\times10^{-7}$\,\solar\,yr$^{-1}$, respectively. We determine the terminal wind velocity assuming $v_{\infty}=2.6v_{\rm esc}$ \citep[][see also \citealp{Barnes2020}]{mcleod_2019}, where $v_{\rm esc}$ is the escape velocity obtained from \citet[][$v_{\rm esc}=\{892,908,923\}$\,\kms]{muijres_2012}. Although our upper limit on the stellar mass represents the limiting case for this scenario, it is worth noting that both observations \citep{Mokiem2007} and simulations \citep{Offner2015} show that the mass loss rates from early-type B stars predicted from models of wind launching \citep[][]{Vink2001} can be underestimated by orders of magnitude \citep[see Figure~3 of][]{Smith2014}. In some cases, the mass loss rates can be as high as the model-predicted mass loss rates of the more massive O-stars considered here (albeit with moderately slower winds).  

Next, we use the mass of the arc to estimate the initial density of the cloud prior to the star's passage, assuming this gas originally filled the volume defined by the radius of the arc. For $M_{\rm arc}\sim2700^{+3000}_{-1400}$\,M$_{\odot}$, we find $\rho_0=3M_{\rm arc}/4\pi R_{\rm arc}^{3}=2.1^{+2.3}_{-1.1}\times10^{-20}$\,g\,cm$^{-3}$, corresponding to a number density $\sim0.9^{+1.0}_{-0.5}\times10^{4}$\,cm$^{-3}$ \citep[which is comparable to the mean density of \brick;][]{federrath_2016, mills_2018}. Finally, we assume $v_{*}=v_{\rm exp}=5.2^{+2.7}_{-1.9}$\,\kms \ and $T=50$\,K (\S~\ref{results:energetics}), such that $c_{s,\rm mol}=0.42$\,\kms, and compute stand-off distances spanning the extremes of this parameter space. The smallest (largest) stand-off distance is set by the upper (lower) limits in the stellar wind properties and the lower (upper) limits in density and $v_{*}$. The range in parameters described above produces stand-off distances of the order 0.01\,pc - 0.1\,pc. The predicted size of the shell is therefore at least an order of magnitude smaller than the observed size of the arc.  

Looking at this another way, for the star to plausibly be an interloper, it must be able to move a distance of order $L=4.7$\,pc within the star's lifetime, $t_{*}$, otherwise it is likely that the star was born right next to the cloud. The maximum stand-off distance (for a fixed mass loss rate and wind speed) is given by the lowest possible relative velocity between the star and the cloud. Assuming a lifetime of $t_{*}\sim20$\,Myr \citep[the limiting case is given by the longest lifetime, and therefore the B1 star;][]{Hurley2000}, this sets a minimum velocity of $v_{\rm min} = L / t_{*}\sim0.2$\,\kms, which in turn gives a maximum standoff distance of $R_{\rm st}=0.8$\,pc (assuming the upper limits in the stellar wind properties and the lower limit in density), which is smaller than what we observe.

In summary, it is difficult to reconcile the fiducial mass and radius estimates of the arc with those predicted assuming that the arc is a swept up shell driven by a stellar wind of a $\approx12\mhyphen20$\,M$_{\odot}$ interloper star moving relative to the cloud. 
Reconciliation may be possible if: i) our assumed mass loss rate and wind velocity are underestimated; ii) both $\rho_0$ and $v_{*}$ are overestimated. Regarding the former scenario, Some of the `field' high-mass stars located within the Galactic Centre are more evolved Wolf-Rayet (WR) stars \citep{mauerhan_2010, dong_2011, Clark2021}. WR stars have powerful stellar winds, with mass loss rates that can be $100\times$ that of O stars. However, they are also more luminous, with Lyman continuum ionising fluxes that are at least an order of magnitude greater than our upper limit derived in \S~\ref{results:hii} \citep[$N_\mathrm{NLyC}>48.6$;][]{Crowther2007}. Therefore it is unlikely that an interloper WR is generating the arc. Regarding the latter scenario, assuming $v_{*}=v_{\rm exp}$, the ambient density would have to be $\sim 3$ orders of magnitude lower than our fiducial value estimated above (since $R_{\rm st}\propto \rho_0^{-1/2}$). This would imply a swept-up mass so small that the arc would be undetectable in dust emission in the current observations. Therefore, a reduction in both $\rho_0$ and $v_{*}$ would be needed to reproduce the observed morphology of the arc. Better mass constraints on the arc would help to conclusively rule out this scenario. As discussed in \S~\ref{results:energetics}, it is not implausible that the mass estimate that we derive for the arc from dust continuum emission is overestimated, particularly if the bulk of that mass is attributed to a spatially overlapping, but unrelated part of the cloud \citep{henshaw_2019}.  

\subsection{Is the arc the result of stellar feedback from in-situ star formation? }\label{discussion:insitu}

An alternative hypothesis to that presented in \S~\ref{discussion:bowshock} is that the arc may be the result of stellar feedback associated with in-situ star formation within \brick. To test that this hypothesis we compare the morphology and dynamics of the arc to analytic prescriptions describing the expansion of H{\sc ii} regions. 

\subsubsection{Thermal expansion of an HII region}\label{discussion:thermal}

The analytic expression for radial expansion of an H{\sc ii} region driven purely by thermal pressure (i.e. with negligible contributions from radiation pressure\footnote{Note that throughout this discussion we neglect radiation pressure from our analysis. Radiation pressure is only important compared to ionised gas pressure when the radius of the H\,{\sc ii} region is below a characteristic radius defined by $R_\mathrm{ch}=0.06 f_\mathrm{trap}^2 S_{49}$\,pc \citep{krumholz_2009}, where $f_\mathrm{trap}$ represents the factor by which the radiation-pressure force is enhanced by trapping of energy within the expanding shell, and $S_{49}$ is the ionising luminosity in units of $10^{49}$\,s$^{-1}$. Taking the upper limit of our range for the ionising luminosity $N_{\rm LyC}=10^{47.9}$\,photons\,s$^{-1}$ (\S~\ref{results:hii}) gives, $R_\mathrm{ch}\approx5\times10^{-3} f_\mathrm{trap}^2$, which is much smaller than the radius of the arc unless $f_\mathrm{trap}>16$. We therefore conclude that radiation pressure is not the likely driving source of the arc. } and stellar winds) is given \citep{spitzer_1978}
\begin{equation}
    R_{\rm Sp}(t)=R_{s}\bigg(1+\frac{7}{4}\frac{c_{s, i}t}{R_{s}}\bigg)^{4/7},
    \label{eq:spitz}
\end{equation}
where $c_{s, i}$ is the sound speed in the ionised gas, $t$ is the age of the H{\sc ii} region, and $R_{s}$ is the Str\"omgren radius. The sound speed in the ionised gas is
\begin{equation}
    c_{s, i}=\sqrt{2.2\frac{k_{B}T_{i}}{\mu m_{\rm H}}},
\end{equation}
where $k_{B}$ is the Boltzmann constant, $T_{i}$ is the temperature of the ionised gas, $\mu$ is the mass per hydrogen nucleus in units of $m_{\rm H}$. The factor of 2.2 arises because there are 2.2 free particles per H nucleus (0.1 He per H, and 1.1 electrons per H; \citealp{krumholz_2017b}). Assuming an ionised gas temperature of $T_{i}=5\times10^{3}$\,K \citep{lang_1997, deharveng_2000, law_2009}, $c_{s, i}\approx8$\,\kms. The Str\"omgren radius is 
\begin{equation}
    R_s = \bigg(\frac{3N_{\rm LyC}\mu^{2}m_{\rm H}^{2}}{4(1.1)\pi\alpha_{B}\rho_{0}^{2}} \bigg)^{1/3},
    \label{Eq:stromgren}
\end{equation}
where we have used the formalism from \citet[their equation 7.24]{krumholz_2017b}. Here, if $\mu=1.4$, the mean mass per hydrogen nucleus in the gas in units of $m_{\rm H}$ and $\rho_{0}$ is the initial density before the photoionizing stars turn on, then $n_{p}=\rho_{0}/\mu m_{\rm H}$ and $n_{e}=1.1\rho_{0}/\mu m_{\rm H}$ with the factor of 1.1 coming from assuming that He is singly ionized and from a ratio of 10 He nuclei per H nucleus. Following \S~\ref{discussion:bowshock}, we present here only the limiting case and assume $N_{\rm LyC}=10^{47.9}$\,photons\,s$^{-1}$. Combining with an initial density $\rho_0=2.1^{+2.3}_{-1.1}\times10^{-20}$\,g\,cm$^{-3}$ (\S~\ref{discussion:bowshock}), the estimated Str\"omgren radius is $R_s\approx0.05^{+0.03}_{-0.02}$\,pc. 

We can use Equation~\ref{eq:spitz} to estimate the time it would take for an H~{\sc ii} region to expand to the observed radius of the arc,
\begin{equation}
    t_{\mathrm{Sp}}=\frac{4}{7}\frac{R_{s}}{c_{s, i}}\bigg[\bigg(\frac{R_{\mathrm{Sp}}}{R_{s}}\bigg)^{7/4}-1\bigg].
\end{equation}
The corresponding velocity with which the H{\sc ii} region expands is given 
\begin{equation}
    v_{\rm Sp}(t) = c_{s, i}\left(1+\frac{7c_{s, i} t}{4R_s}\right)^{-3/7}.
\end{equation}
Equating $R_{\mathrm{Sp}}=R_{\rm arc}$, we find that the estimated age of the H\,{\sc ii} region would be $t_{\mathrm{Sp}}=1.0^{+0.4}_{-0.3}\times10^{6}$\,yr. After $\sim1$\,Myr, the corresponding expansion velocity is expected to be $v_\mathrm{Sp}=0.7^{+0.3}_{-0.2}$\,\kms.

In Figure~\ref{Figure:radii}, we show the time evolution of both the radial expansion (top panel) and the velocity (bottom panel) predicted by the \citet[blue dotted lines][]{spitzer_1978} model. The two curves (blue dotted lines) represent the upper and lower limits on the radial evolution. These limits come from the upper and lower limits on the mass and therefore density (see Equation~\ref{Eq:stromgren}). The shaded region therefore represents the range of parameter space spanned by our estimates of the physical properties. We also include in this figure the model described in \citet{hosokawa_2006}, which also describes thermal expansion but with a slight modification (red dot-dashed lines):
\begin{equation}
    R_{\rm H\&I}(t)=R_{s}\bigg(1+\frac{7}{4}\sqrt{\frac{4}{3}}\frac{c_{s, i}t}{R_{s}}\bigg)^{4/7}.
\end{equation}
Using the \citet{hosokawa_2006} model, the predicted age and velocity of the H~{\sc ii} region are $t_{\mathrm{H\&I}}=0.9^{+0.4}_{-0.3}\times10^{6}$\,yr and $v_\mathrm{H\&I}=~0.8^{+0.3}_{-0.2}$\,\kms, respectively. 

As an H~\textsc{ii} region expands, the photoionised gas in its interior exerts a pressure force and delivers outward radial momentum and kinetic energy to the swept-up shell. \citet[their equation 7.36]{krumholz_2017b} shows that the momentum delivered to the ambient medium, assuming a spherical H~\textsc{ii} region and an ionised gas temperature of $10^4$ K, is
\begin{equation}
    \begin{aligned}
        p = 1.5\times10^5\Bigg[\frac{n_{\rm H}}{10^{2}\,\mathrm{cm}^{-3}}\Bigg]^{-1/7}\Bigg[\frac{N_{\rm ly}}{10^{49}\,\mathrm{s}^{-1}}\Bigg]^{4/7}\Bigg[\frac{t}{10^{6}\,\mathrm{yr}}\Bigg]^{9/7}\Bigg[\frac{T_{e}}{10^{4}\,\mathrm{K}}\Bigg]^{-8/7} \times \\
        \mathrm{M}_{\odot}\,\mathrm{km}\,\mathrm{s}^{-1},
    \end{aligned}
\end{equation}
where $n_{\rm H}$ is the number density of H nuclei in the ambient medium into which the H~\textsc{ii} region is expanding, and $t$ is its age. The expected kinetic energy of the swept-up shell is \citep[equation 7.35]{krumholz_2017b}
\begin{equation}
    \begin{aligned}
        E = 8.1\times 10^{47} \bigg[\frac{n_{\rm H}}{10^{2}\,\mathrm{cm}^{-3}}\bigg]^{-10/7}\bigg[\frac{N_{\rm ly}}{10^{49}\,\mathrm{s}^{-1}}\bigg]^{5/7}\Bigg[\frac{t}{10^{6}\,\mathrm{yr}}\bigg]^{6/7}\Bigg[\frac{T_{e}}{10^{4}\,\mathrm{K}}\Bigg]^{10/7}\times\\
        \,\mathrm{erg}.
    \end{aligned}
\end{equation}
We can use the predicted age of the H~{\sc ii} region therefore, to evaluate the momentum and energy at $t=t_\mathrm{Sp}$. Using our fiducial estimates $N_{\rm ly}=10^{47.9}$\,s$^{-1}$ (\S~\ref{results:hii}), $\rho_0=2.1^{+2.3}_{-1.1}\times10^{-20}$\,g\,cm$^{-3}$ ($n_{\rm H}\sim0.9^{+1.0}_{-0.5}\times10^{4}$\,cm$^{-3}$), and $T_{e}=5\times10^{3}$\,K, we find $p=3.4^{+2.7}_{-1.5}\times10^{4}$\,\solar\,\kms and $0.6^{+1.7}_{-0.5}\times10^{44}$\,erg, respectively.

The above predictions are in considerable tension with the observations. The predicted age of the H{\sc ii} region, implied by the radius of the arc, is almost an order of magnitude greater than the arc's estimated dynamical age (which assumes that the expansion velocity has been constant over this time; \S~\ref{results:toy}). Although the predicted momentum only differs from our measured value by a factor of $2-3$, the predicted velocity and energy show considerably more tension with the measured quantities, differing by factors of $\sim$1 and 4 orders of magnitude, respectively. Given that this calculation uses our upper limit on the estimated Lyman continuum ionising flux, and therefore represents a best case scenario for this hypothesis, we are able to rule out thermal expansion of an H\,{\sc ii} region as the possible driving source of the arc.

\subsubsection{A wind-blown bubble}\label{discussion:wind}

The analysis presented in the previous section indicates that there must be a significant source of energy on top of that provided by the thermal pressure of photoionised gas. One possibility is that this energy is provided by the stellar wind. In the following, we explore the possibility that the arc represents the dense, partial shell that surrounds a bubble driven by a stellar wind from a high-mass star. 
The time evolution of radial expansion of a bubble driven by stellar winds can be expressed \citep{weaver_1977},
\begin{equation}
R_\mathrm{W}(t) = \alpha\bigg( \frac{L_\mathrm{wind}}{\rho_0}\bigg)^{1/5}t^{3/5},
\label{eq:rw}
\end{equation}
where $\alpha=[125/154(\pi)]^{1/5}$ \citep{tielens_2005, Lancaster2021a}, $L_{\rm wind}$ is the mechanical wind luminosity, $L_{\rm wind}=0.5 \dot{M}v_{\infty}^{2}$, $\rho_{0}$ is the ambient density (estimated in \S~\ref{discussion:bowshock}).

The \citet{weaver_1977} solution assumes that the wind gas is adiabatic and trapped, so it applies to a bubble that is completely closed and has no cooling. As soon as gas breaks out, or there is significant mixing between hot and cold gas that leads to cooling, the rate of expansion will drop below the \citet{weaver_1977} solution \citep{McKee1984,MacLow1988,Lancaster2021a}. \citet{MacLow1988} relaxed the condition that the wind gas is adiabiatic and included radiative cooling from the interior of the bubble. At early times, the expansion follows the analytic \citet{weaver_1977} solution. At later times, some of the internal energy is radiated away and the expansion rate slows. The numerical solution of \citet{MacLow1988} grows at a rate close to $t^{1/2}$, such that we can write
\begin{equation}
    R_\mathrm{W_c}(t)=R_\mathrm{cool}\bigg(\frac{t}{t_\mathrm{cool}}\bigg)^{1/2},
\end{equation}
where $R_\mathrm{cool}=R_\mathrm{W}$, given by Equation~\ref{eq:rw}, is the radius of the bubble at a time $t=t_\mathrm{cool}$, where $t_\mathrm{cool}$ is the time at which radiative cooling becomes significant. Using this expression, we can estimate the time it would take for a wind-blown bubble to expand its current size assuming this time is $>t_{\rm cool}$:
\begin{equation}
\label{eq:tw}
t_\mathrm{W_c} = \bigg(\frac{R_\mathrm{W_c}}{\alpha}\bigg)^{2}\bigg( \frac{L_\mathrm{wind}}{\rho_0}\bigg)^{-2/5}t_\mathrm{cool}^{-1/5}.
\end{equation}
The corresponding expansion velocity, momentum in the shell, and kinetic energy of the shell are
\begin{equation}
v_\mathrm{W_c} = \frac{1}{2}\alpha \bigg( \frac{L_\mathrm{wind}}{\rho_0}\bigg)^{1/5}t_\mathrm{W_c}^{-1/2}t_\mathrm{cool}^{1/10},
\label{eq:vw}
\end{equation}
\begin{equation} 
p_\mathrm{W_c} = M_\mathrm{arc}v_\mathrm{W_c}=\frac{2\pi}{3}\alpha^{4}(L_\mathrm{wind}^{4}\rho_{0}t_\mathrm{cool}^{2})^{1/5}t_\mathrm{W_c},
\end{equation}
and
\begin{equation}
    E_\mathrm{W_c}= \frac{1}{2}M_\mathrm{arc}v^{2}_\mathrm{W_c}= \frac{125}{462}L_\mathrm{wind}t_\mathrm{W_c}^{1/2}t_\mathrm{cool}^{1/2},
\end{equation}
respectively.

The cooling time can be expressed \citep{MacLow1988,Chevance2020}
\begin{equation}
    t_\mathrm{cool}\approx3000\bigg(\frac{Z}{\mathrm{Z_{\odot}}}\bigg)^{-35/22}\bigg(\frac{L_\mathrm{wind}}{10^{35}\,\mathrm{erg\,s}^{-1}}\bigg)^{3/11}\bigg(\frac{n_\mathrm{H}}{10^{4}\,\mathrm{cm}^{-3}}\bigg)^{-8/11}\,\mathrm{yr},
\end{equation}
where $Z$ is the metallicity. To estimate the cooling time, we must therefore estimate the mechanical wind luminosity. As discussed in \S~\ref{discussion:bowshock}, both observations \citep{Mokiem2007} and simulations \citep{Offner2015} show that the mass loss rates from early-type B stars predicted from the models of wind launching considered here \citep[][]{Vink2001} can be underestimated by orders of magnitude. In the following, we therefore use the mass loss rates and terminal wind velocities derived for O stars of spectral type O9.5, O9, O8.5 in \S~\ref{discussion:bowshock}, under the assumption that these provide the limiting case for this scenario. We therefore estimate the range in mechanical wind luminosity that spans this parameter space, finding $L_{\rm wind}=0.4-2.2\times10^{35}$\,erg\,s$^{-1}$ \citep[note that in some cases empirically derived mechanical wind luminosities from early type B stars can actually exceed this range; ][]{Mokiem2007}. Inserting numerical values we derive a range of cooling times $t_\mathrm{cool}=1500\mhyphen2200$\,yr, where the lower limit is given by our lower limit on the mechanical wind luminosity and the upper limit on the cloud density at solar metallicity (the upper limit is given by the opposite at twice solar metallicity). Due to the considerable ambient density of G0.253+0.016, the corresponding cooling time is much shorter than that inferred under the typical conditions found in galaxy discs \citep{MacLow1988, Chevance2020}. Using Equation~\ref{eq:rw}, the corresponding size of the wind blown bubble at time $t=t_\mathrm{cool}$ is therefore $R_\mathrm{cool}=0.05\mhyphen0.12$\,pc.

\begin{figure}
\begin{center}
\includegraphics[trim = 10mm 13mm 0mm 5mm, clip, width = 0.48\textwidth]{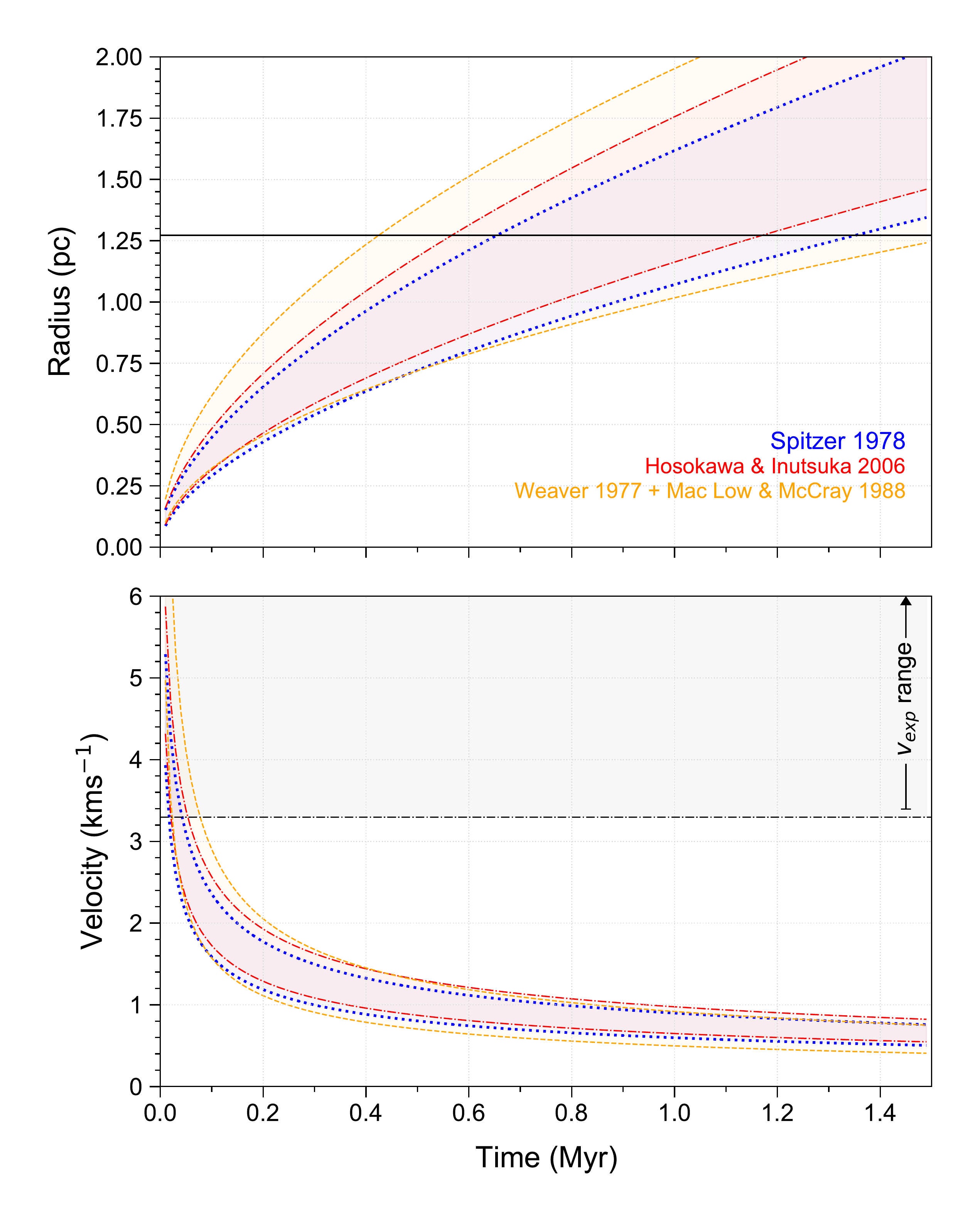}
\end{center}
\caption{The top panel shows analytic predictions for the time evolution of the radii of expanding H{\sc ii} regions from various models (see text for details). The blue dotted curve indicates expansion driven by the thermal pressure of photoionised gas \citet{spitzer_1978}. The red dot-dashed curve is the same but with a slight modification from \citet{hosokawa_2006}. The orange dashed curves describe the radial expansion driven by stellar winds for stars of different spectral types consistent with our measurement of $N_{\rm LyC}$ (\citealp{weaver_1977, MacLow1988}). The horizontal black line represents the radius of the arc $R_{\rm arc}=1.3$\,pc. The bottom panels show the corresponding time evolution of the expansion velocity. The black shaded region indicates the range of expansion velocity derived from the different methods presented in \S~\ref{results:toy} (note that this has been truncated for clarity, as indicated by the black arrow). The horizontal dot-dashed line reflects the lower limit of the expansion velocity estimates shown in Figure~\ref{Figure:pvarc}.  }
\label{Figure:radii}
\end{figure}

In the top panel of Figure~\ref{Figure:radii}, we show curves corresponding to the time evolution of wind-blown bubbles that represent the extremes of the parameter space described above (orange dashed lines). The model in which the shell swept up by the wind-blown bubble expands most quickly (slowly) is derived from our upper (lower) limits on the stellar mass and metallicity, but the lower (upper) limit on density. The corresponding evolution in the expansion velocity is shown in the bottom panel. Equating $R_\mathrm{W_c}=R_{\rm arc}$, for $M/{\rm M}_{\odot}=19.82$, $\dot M(Z/{\rm Z}_{\odot}=2)$, and $n_{\rm H}\sim0.4\times10^{4}$\,cm$^{-3}$, we derive an age of $t_\mathrm{W_c}=0.4\times10^{6}$\,yr, an expansion velocity of $v_\mathrm{W_c}=1.5$\,\kms, a momentum $p_\mathrm{W_c}=0.2\times10^{4}$\,\solar\,\kms, and an energy $E_\mathrm{W_c}=0.6\times10^{47}$\,erg. The same calculation for $M/{\rm M}_{\odot}=16.46$, $\dot M(Z/{\rm Z}_{\odot}=1)$, and $n_{H}\sim1.9\times10^{4}$\,cm$^{-3}$ yields $t_\mathrm{W_c}=1.6\times10^{6}$\,yr, $v_\mathrm{W_c}=0.4$\,\kms, $p_\mathrm{W_c}=0.2\times10^{4}$\,\solar\,\kms, and $E_\mathrm{W_c}=0.2\times10^{47}$\,erg.

For the $M=16.46$\,M$_{\odot}$ star, the expansion velocity and momentum are an order of magnitude below the values estimated from the observations, but the predicted energy is lower by $>$2 orders of magnitude. In the case of the $M=19.82$\,M$_{\odot}$ star, the predicted momentum and energy are comparable to within a factor of $<2$ to the measured values, while the predicted expansion velocity is lower by a factor of $\sim3.5$ compared to our fiducial estimate of $5.2$\,\kms.\footnote{Note that for the $M=16.46$\,M$_{\odot}$ star we compare our predicted values to our measured upper limit on density and momentum (worst case scenario) and for the $M=19.82$\,M$_{\odot}$ star the reverse (best case scenario).} While the agreement remains imperfect, this analysis demonstrates that the arc could plausibly represent a dense, partial shell surrounding a bubble driven by a stellar wind. The factor of $\sim$a few discrepancy may be explained by the fact that each of the discussions above consider a single feedback mechanism acting in isolation when in reality different mechanisms may act in concert \citep{Draine2011b, Martinez-Gonzalez2014, Yeh2013, Mackey2015}. A full prescription of the different feedback mechanisms is beyond the scope of the present study and will require detailed modelling tailored to the conditions found in G0.253+0.016 and, more generally, the extreme environment of the CMZ. 

\subsection{Is a wind-blown bubble the most likely scenario?}\label{discussion:context}

The analysis presented in the previous sections leads us to conclude the following:
\begin{enumerate}
    \item the arc is plausibly the result of stellar feedback.
    \item the estimated density and morphology of the arc are difficult to reconcile with a scenario in which the arc is a bow-shock swept up by the wind of an interloper star.
    \item the thermal pressure of photoionised gas alone is unable to reproduce the estimated dynamics and energetics of the arc.
    \item the arc may represent a dense, partial shell surrounding a bubble driven by the wind from a high-mass star.
\end{enumerate}

The importance of winds from high-mass stars as a feedback mechanism is under recently revived debate. Numerical simulations have had a consensus for some time that generally photoionisation dominates over winds \citep[][]{Dale2013, Rathjen2021, Geen2021}. Despite this, there are several sources with morphology and dynamics which appear to be consistent with those expected for wind-blown bubbles. RCW\,120 has been recently described as being a wind-blown bubble driven by a O8V star moving relative to the ambient cloud material by $<4$\,km\,s$^{-1}$, with further evidence to suggest that star formation may have been triggered within the swept up shell \citep{Luisi2021}. Similarly, \citet{Pabst2019, Pabst2020} recently concluded that the bubble of the Orion Nebula is predominantly driven by the mechanical energy input of the strong stellar wind from the O7V star $\theta^{1}$ Orionis C \citep[see also][]{Gudel2008}, based on the simple analytic model of \citet{weaver_1977}. 

This latter interpretation however, faces many challenges. As described in \S~\ref{discussion:wind}, the \citet{weaver_1977} solution assumes that the wind gas is adiabatic and trapped. As soon as the gas cools, the expansion speed will drop below the \citet{weaver_1977} solution. The recent work of \citet{Lancaster2021a,Lancaster2021b} demonstrates that turbulence-driven inhomogeneity in the structure of the material surrounding the wind-driven bubbles may strongly affect the impact of the mechanical energy of the wind. The cooling induced by turbulent mixing in the absence of magnetic fields leads to order of magnitude differences in the expansion velocity and imparted momentum compared to those derived in the classical \citet{weaver_1977} solution, although there is evidence that magnetic fields at least partly mitigate this effect \citep[e.g.,][]{Gentry19a}. Indeed, the recent numerical simulations of \citet{Rosen2021} also show that wind bubbles blown by individual high-mass stars do not experience efficient mixing in the presence of magnetic fields \citep[][estimate a total magnetic field strength of $5.4\,\pm\,0.5$\,mG in \brick]{Pillai2015}. The magnetic field provides a confining and stabilising effect and suppresses the development of instabilities that otherwise lead to effective mixing and cooling \citep{Lancaster2021a,Lancaster2021b}. It is also worth noting that direct measurements of the X-ray luminosities of wind-blown bubbles are inconsistent with the \citet{weaver_1977} model, and require substantial loss of energy via either turbulent mixing or bulk escape of hot material \citep{Harper-Clark2009, Rosen2014}. It may therefore simply be the case that the high velocity C~\textsc{ii} emission observed by \citet{Pabst2019, Pabst2020} is tracing material from a wind that is escaping along low-density channels in the bubble, rather than driving feedback globally in the region \citep{Haid2018}. 

In the Galactic Centre, a number of molecular shell candidates have been identified \citep{Martin-Pintado1999, Oka2001, Butterfield2018, Tsujimoto2018,Tsujimoto2021}. The kinetic energy estimated for many of these shells has led to speculation that they are the result of (potentially multiple) supernova explosions \citep[e.g.][]{Tsujimoto2018}. However, those identified in Sgr\,B2 by \citet{Martin-Pintado1999} share many of the properties displayed by the arc in \brick. \citet{Martin-Pintado1999} identify a series of $\sim1-2$\,pc shells and arcs detected in emission from the (3,3) and (4,4) lines of NH$_3$. (Recall that the arc in \brick \ is also prominent in these lines -- \citealp{mills_2015}.) They conclude that the shells are expanding with velocities $6-10$\,\kms \ and have an associated kinetic energy of the order $10^{48}$\,erg, very similar to the quantities derived for the arc in \brick \ and considerably smaller than typical energies of $\sim10^{51}$\,erg associated with supernova-driven shells. The authors speculate that the shells in Sgr\,B2 are produced by the wind-blown bubbles generated by high-mass stars and describe how the shocks generated by the expansion heat the surrounding gas, further arguing that the expanding shells may have even triggered further star formation within Sgr\,B2's envelope. 

The arc located in \brick \ provides an interesting new addition to this puzzle. First, the associated radio continuum emission is extended, unlike the compact H\,{\sc ii} regions driven by O-type stars in other clouds in the Galactic Centre \citep[e.g., Sgr A A-D and H][]{Goss_1985,Zhao_1993,Mills2011,Hankins2019}. One possible explanation for this may be because the source driving the arc is less embedded, having formed at the edge of the cloud and excavated a cavity. Second, the morphology, dynamics, and energetics of the arc show reasonable (to within a factor of a few) agreement with a modified form of the \citet{weaver_1977} solution that accounts for cooling within the bubble interior \citep[][]{MacLow1988}, but differs from that in Orion \citep{Pabst2019, Pabst2020} in that it is identified using a molecular (rather than ionised gas) tracer. It is certainly possible that local environmental conditions in the Galactic Centre may help winds to play an important role. In high-density environments, winds may stay contained within the shell longer leading to more prolonged expansion \citep{Barnes2020}. Hence we are left with three possibilities: i) winds are not the key feedback driving mechanism and some other explanation is required to explain the origin of the arc; ii) winds are more important for driving feedback than otherwise expected, in such a way that simulations, and the interpretation of observations of winds (e.g.\ in X-rays) are incorrect; iii) winds are less important under normal conditions, but may be more important under the extreme conditions (e.g., high-density, high-metallicity, strong magnetic fields) in the Galactic Centre \citep[e.g.][]{Martin-Pintado1999, Barnes2020}.

\subsection{Has \brick \ already formed a star cluster?}\label{discussion:cluster}

\begin{figure*}
\begin{center}
\includegraphics[trim = 0mm 5mm 0mm 5mm, clip, width = 0.85\textwidth]{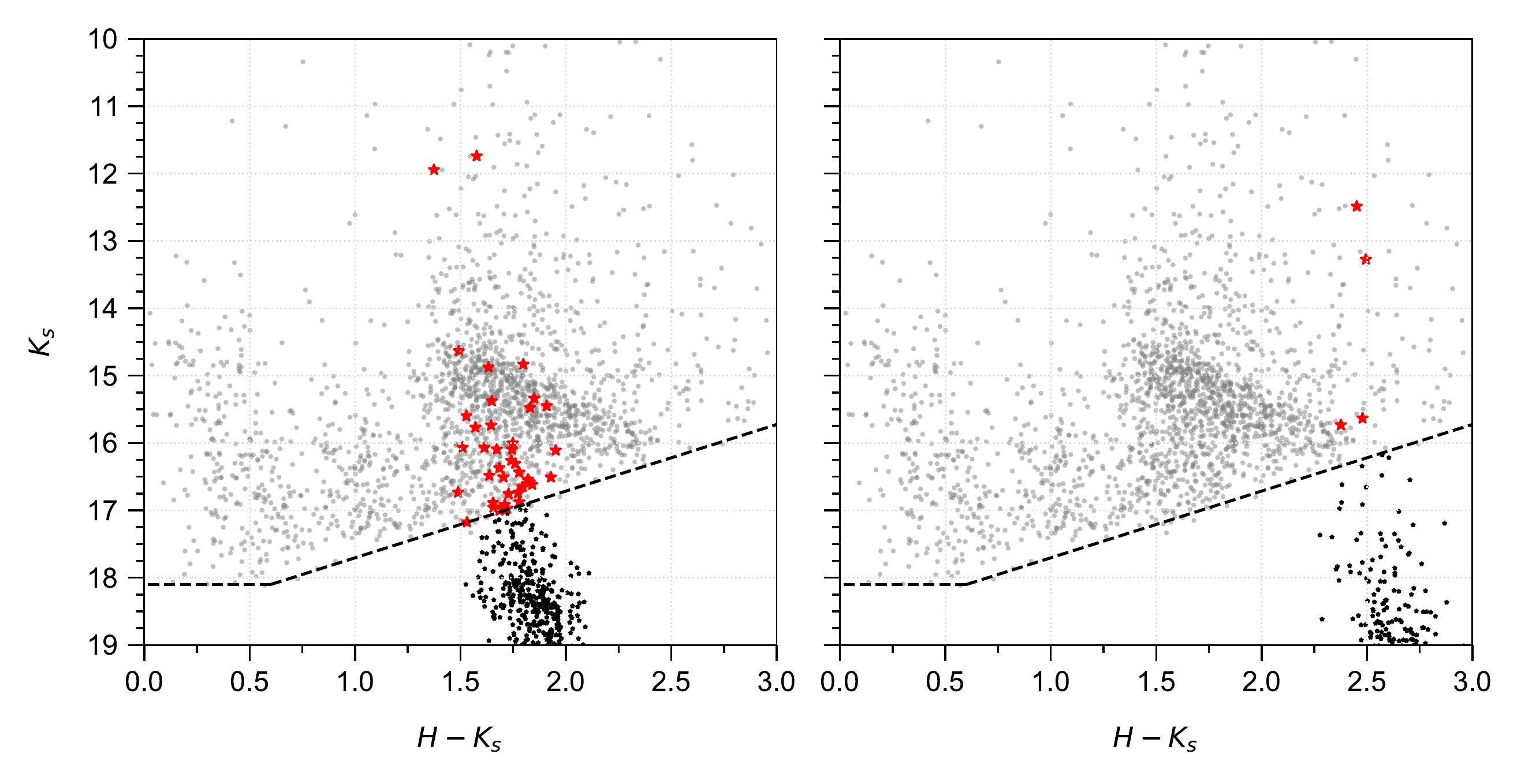}
\end{center}
\caption{Colour-magnitude diagram $K_s$ versus $H-K_s$ corresponding to the region containing \brick \ \citep[see Fig. 2 in][]{Nogueras-Lara2021a}. Grey dots represent real stars from the GNS survey \citep{Nogueras-Lara2018,Nogueras-Lara2019}. The red and black stars correspond to the synthetic stellar population of a young (0.5\,Myr) cluster of mass 500\,\solar. The red and black stars detections and non-detections from the synthetic population, respectively, considering the detection limit of the data (black dashed line). The left and right panels assume an extinction of $A_{K_s} = 2$ and 3\,mag, respectively, with the latter indicating fewer detections of cluster members.  } 
\label{Figure:cluster}
\end{figure*}

In this section we address the elephant in the room, namely that if the arc is the result of a wind-blown bubble generated by a high-mass star, then \emph{where is the star?} The short $\sim760\mhyphen7000$\,yr recombination time estimated in \S~\ref{results:hii} implies that the source of the ionising radiation must still reside within the cavity enclosed by the arc. If the star has formed in situ, as implied by the wind-blown bubble scenario, then the immediate implication is that \brick \ is perhaps not as quiescent as is commonly accepted. High-mass stars rarely (if at all) form in isolation \citep{deWit2004,deWit2005}. Though isolated high-mass stars have been identified throughout the Galactic Centre \citep{mauerhan_2010, dong_2011, Clark2021}, the cluster formation efficiency in CMZ clouds may be as high as $\sim30\mhyphen40\%$ \citep{ginsburg_2018b}.

Assuming the high-mass star forms as part of a star cluster, we can estimate the mass of the parent cluster and address the question of whether or not we would be likely to detect such a cluster towards \brick. Again here, we consider only the O star scenario, since this presents the best case scenario for detectability. To estimate the mass of the parent star cluster, we simulate samples of star clusters for a range of cluster masses, generating $n=10000$ clusters of each mass, assuming a standard stellar IMF \citep{kroupa_2001}. For each cluster we determine the mass of its highest mass star, comparing the peak of the distribution to the $16\mhyphen20$\,\solar \ relevant for stars of spectral type consistent with our upper limit of the Lyman continuum ionising flux, $N_{\rm LyC}$ \citep{Martins2005}. We find that cluster masses of the order $400\mhyphen700$\,\solar \ are typical for those in which the most massive star is $\sim16\mhyphen20$\,\solar. 

As an independent estimate of the potential cluster mass, we follow the method outlined in \citet{barnes_2017}. To do this, we first estimate the bolometric luminosity from infrared luminosity maps of the CMZ using \emph{Spitzer} and \emph{Herschel} observations. \citet{barnes_2017} assume that all the emission from the embedded stellar population within a molecular cloud is reprocessed by the surrounding dust and re-emitted. Under this assumption the total infrared luminosity directly corresponds to the bolometric luminosity produced by the embedded population. We apply this method to the arc by estimating the total bolometric luminosity within the region defined in Figure~\ref{Figure:arc_SS}, for which we find $L_{\rm bol}\sim1.2\times10^{5}$\,L$_{\odot}$. We can convert this bolometric luminosity to a stellar mass by assuming that the highest mass star within the cluster dominates the luminosity. To do this, we use the bolometric luminosity-to-mass conversions presented by \citet{davies_2011}. For $L_{\rm bol}\sim1.2\times10^{5}$\,L$_{\odot}$ we find $M_{*}\sim31$\,M$_{\odot}$. Repeating the same experiment as before, we find that a cluster mass of the order $\sim1000$\,\solar \ is typical for those in which the most massive star is $\sim31$\,\solar. Given the uncertainty in equating the total infrared luminosity to bolometric luminosity, this should be interpreted as a strict upper limit on the total mass of the embedded stellar population (see \citealp{barnes_2017} for further details). Although the absolute values should be taken with caution, this analysis suggests that the independent measures of radio continuum emission and the total infrared luminosity are consistent with the presence of a (moderately) high-mass star.

To address whether we would be expected to detect such a star cluster in currently available data, we use the GALACTICNUCLEUS (GNS) catalogue. The GNS is a high-angular resolution  ($\sim0\farcs2$) $JHK_s$ survey of the Galactic Centre \citep{Nogueras-Lara2018, Nogueras-Lara2019}, that partially covers \brick. We build a synthetic young cluster with a total mass of $500\,M_\odot$, using PARSEC evolutionary tracks \citep{Bressan2012, Chen2014, Chen2015, Tang2014, Marigo2017, Pastorelli2019,Pastorelli2020} to obtain $H$ and $K_s$ photometry. We assume twice solar metallicity \citep{Feldmeier-Krause2017, Schultheis2019, Schultheis2021} and a standard IMF \citep{kroupa_2001} and create five different models with different ages (0.5, 0.7, 1, and 5\,Myr). To redden the data, we test three different scenarios using average extinctions $A_{Ks}=2,2.5,3.0$\,mag. We redden the synthetic data randomly, choosing the extinction value for each star from a Gaussian distribution centred on the average extinctions with a typical standard deviation of $\sim0.1$\,mag \citep{Nogueras-Lara2020}. We randomly simulate the photometric uncertainties for each star assuming a Gaussian distribution for each band, with a standard deviation of 0.05\,mag corresponding to the expected uncertainty for the GNS data \citep{Nogueras-Lara2021}. Finally, we place the stellar population at the Galactic Centre distance using a distance modulus of 14.52 \citep{Nogueras-Lara2021a}. 

We plot the simulated stellar populations on the colour-magnitude diagram (CMD) $K_s$ versus $H-K_s$ towards \brick \ \citep[Figure~\ref{Figure:cluster}][]{Nogueras-Lara2021a}. Using the limitations of the real GNS data, we identify which of the cluster stars may be detected. Assuming the lowest extinction ($A_{K_s}=2.0$\,mag), $\lesssim 40$ stars can be detected for each of the different ages tested and this decreases with increasing cluster age. The most favourable case, in terms of detection, is the youngest cluster age considered (0.5 Myr; Figure~\ref{Figure:cluster}). Given the stellar background in the CMD, the differential reddening, and the very low number of potentially observed stars belonging to the young cluster, we conclude that a direct detection using the CMD would be unlikely. Moreover, the assumed extinction of $A_{K_s}=2.0$\,mag corresponds to the value obtained by \citet{Nogueras-Lara2021a} using red clump stars \citep[e.g.][]{Girardi2016} for the region containing \brick. This is the best case scenario for detection and is equivalent to the cluster being situated in the foreground of the cloud. Assuming a larger extinction of $A_{K_s}=3.0$\,mag, we obtain even fewer detections of the cluster members (Figure~\ref{Figure:cluster}). 

Finally, we also check whether the cluster could be detected due to stellar overdensities in the NIR images. We use the $K_s$ band, where the extinction is lowest, and compute the stellar density using the GNS data corresponding \brick. We divide the observed region into small sub-regions of 1\,pc$^2$ to compute the number of stars detected in $K_s$. Averaging over all the sub-regions, we find a mean stellar surface density of $\sim 180\,\pm\,90$\,pc$^{-2}$, where the uncertainty corresponds to the standard deviation of the measurement. Considering the most favourable case of a cluster stellar population of 0.5\,Myr, an extinction of $A_{K_{s}}=2.5$\,mag, and assuming that the cluster extends to a radius of $\sim0.5$\,pc (comparable to the Arches, \citealp{hosek_2015}), the expected over-density is $\sim80$\,pc$^{-2}$ indicating that the cluster would not easily be detected by its stellar density.

In summary, we conclude that the high-extinction and stellar crowding towards \brick \ is more than capable of hampering the detection of a $500$\,\solar \ star cluster in currently available NIR data. Moreover, we stress that the above assumes best case scenario for detection. To detect such a cluster, longer integration time NIR observations would be needed to detect fainter cluster members. However, this may not help if the cluster were deeply embedded within the cloud or behind the main column. The discussions presented in \S~\ref{discussion:context} and here clearly call for further observations to resolve any ambiguity that remains surrounding the possible origins of the arc. Future high-sensitivity observations with other facilities, such as the James Webb Space Telescope (JWST), will likely reveal the true star formation activity of \brick. 

\subsection{What is the implied star formation rate?}

\citet{barnes_2017} provide an upper limit of the total stellar mass of newly formed stars within the Brick of $>$\,2000\,M$_{\odot}$ from a measurement of the total infrared emission. These authors estimate a star formation rate of $<$\,0.007\,M$_{\odot}$yr$^{-1}$ based on this total stellar mass and a star formation timescale based on inferences about the orbit of the cloud ($t_\mathrm{SF}=0.3$\,Myr). \citet{kauffmann_2017a}, on the other hand, estimate an upper limit of $\sim$\,800\,M$_{\odot}$ based on the absence of any radio or maser emission sources. These authors used a timescale based on a statistical approach based on the number of observed H\,{\sc ii} region and masers within the CMZ ($t_\mathrm{SF}=1.1$\,Myr), and determined a star formation rate of $<$\,0.0008\,M$_{\odot}$yr$^{-1}$. Based on the observed bounds of our derived $N_{LyC}$ values, we estimate here the associated star formation rate of a $12\mhyphen20$\,M$_{\odot}$ star (section\,\ref{results:hii}), under the assumption that this implies the presence of a $\sim$500\,\solar \ cluster (given a standard \citealt{kroupa_2001} IMF). Assuming that the cluster has an age $t_\mathrm{SF}=0.4 \mhyphen 1.6$\,Myr (see \S~\ref{discussion:wind}), the associated star formation rate is in the range $0.0003 \mhyphen 0.0013$\,M$_{\odot}$yr$^{-1}$. The star formation rates are highly dependent on the assumed timescales over which they are inferred. Nonetheless, our estimates based on the presence of a single B1-O8.5 star are broadly consistent with the low star formation rates measured within the literature.

\section{Summary \& Conclusions}\label{conclusions}

In this paper, we have built on the analysis presented in \citet{henshaw_2019}, combining ALMA and VLA observations to determine the origin of the arcuate structure identified within \brick. We find evidence for an expanding bubble associated with ionised gas emission. Our main conclusions are summarised below.

Using the kinematic decomposition presented in \citet{henshaw_2019}, we find that morphology of the arc can be described using a simple tilted ring model. The ring is centred on $\{l,b\}=\{0\fdg248,\,0\fdg018\}$ and has a radius of $R_{\rm arc}=1.3$\,pc. The azimuthal velocity pattern observed along the crest of the arc is broadly consistent with that expected for an expanding incomplete shell. Using our model geometry, we derive an expansion velocity of $v_{\rm exp}=5.2^{+2.7}_{-1.9}$\,\kms. From this information we infer that the dynamical age of the arc is $t_{\rm dyn}\approx2.4^{+0.8}_{-1.4}\times10^{5}$\,yr (assuming a constant expansion velocity). Using dust continuum observations we determine the mass off the arc to be $M_{\rm arc}\sim2700^{+3000}_{-1400}$\,M$_{\odot}$. Combining with the derived expansion velocity, we measure the kinetic energy and momentum of the arc to be $E_{\rm arc}\sim0.7^{+2.8}_{-0.6}\times10^{48}$\,erg and $p_{\rm arc}\sim1.4^{+3.1}_{-1.0}\times10^{4}$\,M$_{\odot}$\kms, respectively. 

Our new radio continuum and radio recombination line (RRL) data reveal that ionised gas fills the arc cavity. The RRL spectrum extracted from the arc cavity peaks at a velocity of $22.0\,\pm\,1.4$\,\kms, consistent to within one standard deviation of the  mean of the arc centroid velocity distribution ($17.6\,\pm\,4.5$\,\kms). The spatial and kinematic agreement between the ionised and molecular gas emission leads us to conclude that the two are likely physically related. To give insight into the type of source required to stimulate this emission, we calculate the Lyman-continuum photon rate, $N_{\rm LyC}=10^{46.0}\mhyphen10^{47.9}$\,photons\,s$^{-1}$. The implied short recombination time of $t_{\rm rec}=760\mhyphen7000$\,yr further suggests that the source of the ionised gas must still be located within the arc cavity. Assuming that the emission is produced by a single zero-age main sequence star, the estimated $N_{LyC}$ is consistent with that expected for a high-mass star of spectral type B1-O8.5, corresponding to a mass of $\approx12\mhyphen20$\,\solar.

We go on to explore the possible origins of the arc and the potential star driving its expansion. We consider two scenarios: i) the arc represents a shell swept up by the wind of an interloper high-mass star; ii) the arc represents a shell swept up by stellar feedback resulting from in-situ star formation. For the former scenario, the CMZ is unique in our Galaxy in that there is a rich population of `field' high-mass stars, and we show that the probability that a high-mass star may be passing through \brick \ at the present time is reasonably high. Nevertheless, we deduce that there does not appear to be a way to reconcile the required ionising continuum with the current mass and radius estimates of the arc under the assumption that the arc represents a bow-shock produced by a slowly moving high-mass star. This size constraint rules out the Arches and Quintuplet clusters as possible sources of any interloper. Given the information currently available to us, we therefore conclude that the arc is plausibly the result of stellar feedback from in-situ star formation. We compare the morphological and dynamical properties of the arc, as well as its estimated kinetic energy and momentum to simple analytical models describing the expansion of H\,{\sc ii} regions, finding that the properties of the arc are consistent to within a factor of a few with those produced by a wind-blown bubble generated by a high-mass stars star. 

The immediate implication of this result is that \brick \ may not be as quiescent as is commonly accepted. Assuming that the high-mass star did not form in isolation, our results could mean that \brick \ has already produced a $\lesssim10^{3}$\,\solar \ cluster, containing at least one high-mass star. We demonstrate that the high-extinction and stellar crowding observed towards \brick \ are more than capable of obscuring such a star cluster from view. Future observations are needed to resolve any residual ambiguity left surrounding the origins of the arc. This is important to establish the true underlying star formation rate of molecular clouds in the CMZ, and to precisely establish the role of stellar feedback in shaping the ISM and regulating the star formation process in an environment which has the highest number of high-mass stars per unit volume in the Galaxy. We suggest that future observations from facilities such as ALMA (to better constrain the mass of the arc) the JWST (to reveal the internal stellar population) will have the sensitivity necessary to confirm or reject this result. 

\section*{Acknowledgements}

We would like to thank the anonymous referee, whose constructive report helped to improve this paper. 
We would like to thank Iskren Georgiev and Héctor Arce for insightful discussions.
MRK acknowledges support from the Alexander von Humboldt Foundation through a Humboldt Research Award, and from the Australian Research Council through its \textit{Discovery Projects} and \textit{Future Fellowship} funding schemes, awards DP190101258 and FT180100375.
JM acknowledges support from a Royal Society-Science Foundation Ireland University Research Fellowship (14/RS-URF/3219, 20/RS-URF-R/3712). 
AG acknowledges support from the National Science Foundation under grant No. 2008101.
TJH is funded by a Royal Society Dorothy Hodgkin Fellowship. 
FN-L gratefully acknowledges support by the Deutsche Forschungsgemeinschaft
(DFG, German Research Foundation) – Project-ID 138713538 – SFB 881 (“The Milky Way System”, subproject B8), and the sponsorship provided by the Federal Ministry for Education and Research of Germany through the Alexander von Humboldt Foundation.
ATB would like to acknowledge funding from the European Research Council (ERC) under the European Union’s Horizon 2020 research and innovation programme (grant agreement No.726384/Empire).
JMDK gratefully acknowledges funding from the Deutsche Forschungsgemeinschaft (DFG, German Research Foundation) through an Emmy Noether Research Group (grant number KR4801/1-1), as well as from the European Research Council (ERC) under the European Union's Horizon 2020 research and innovation programme via the ERC Starting Grant MUSTANG (grant agreement number 714907).
HB acknowledges support from the European Research Council under the Horizon 2020 Framework Program via the ERC Consolidator Grant CSF-648505.  HB also acknowledges support from the Deutsche Forschungsgemeinschaft in the Collaborative Research Center SFB 881 - Project-ID 138713538 - “The Milky Way System” (subproject B1).
DW and CB acknowledge support from the National Science Foundation under Award No. 1816715.

\section*{Data Availability}

The data underlying this article will be shared on reasonable request to the corresponding author.




\bibliographystyle{mnras}
\bibliography{references} 

\begin{thebibliography}{}
\makeatletter
\relax
\def\mn@urlcharsother{\let\do\@makeother \do\$\do\&\do\#\do\^\do\_\do\%\do\~}
\def\mn@doi{\begingroup\mn@urlcharsother \@ifnextchar [ {\mn@doi@}
  {\mn@doi@[]}}
\def\mn@doi@[#1]#2{\def\@tempa{#1}\ifx\@tempa\@empty \href
  {http://dx.doi.org/#2} {doi:#2}\else \href {http://dx.doi.org/#2} {#1}\fi
  \endgroup}
\def\mn@eprint#1#2{\mn@eprint@#1:#2::\@nil}
\def\mn@eprint@arXiv#1{\href {http://arxiv.org/abs/#1} {{\tt arXiv:#1}}}
\def\mn@eprint@dblp#1{\href {http://dblp.uni-trier.de/rec/bibtex/#1.xml}
  {dblp:#1}}
\def\mn@eprint@#1:#2:#3:#4\@nil{\def\@tempa {#1}\def\@tempb {#2}\def\@tempc
  {#3}\ifx \@tempc \@empty \let \@tempc \@tempb \let \@tempb \@tempa \fi \ifx
  \@tempb \@empty \def\@tempb {arXiv}\fi \@ifundefined
  {mn@eprint@\@tempb}{\@tempb:\@tempc}{\expandafter \expandafter \csname
  mn@eprint@\@tempb\endcsname \expandafter{\@tempc}}}

\bibitem[\protect\citeauthoryear{{Armentrout}, {Anderson}, {Balser}, {Bania},
  {Dame}  \& {Wenger}}{{Armentrout} et~al.}{2017}]{Armentrout2017}
{Armentrout} W.~P.,  {Anderson} L.~D.,  {Balser} D.~S.,  {Bania} T.~M.,  {Dame}
  T.~M.,   {Wenger} T.~V.,  2017, \mn@doi [\apj] {10.3847/1538-4357/aa71a1},
  \href {https://ui.adsabs.harvard.edu/abs/2017ApJ...841..121A} {841, 121}

\bibitem[\protect\citeauthoryear{{Armillotta}, {Krumholz}, {Di Teodoro}  \&
  {McClure-Griffiths}}{{Armillotta} et~al.}{2019}]{Armillotta2019}
{Armillotta} L.,  {Krumholz} M.~R.,  {Di Teodoro} E.~M.,   {McClure-Griffiths}
  N.~M.,  2019, \mn@doi [\mnras] {10.1093/mnras/stz2880}, \href
  {https://ui.adsabs.harvard.edu/abs/2019MNRAS.490.4401A} {490, 4401}

\bibitem[\protect\citeauthoryear{{Arthur} \& {Hoare}}{{Arthur} \&
  {Hoare}}{2006}]{arthur_2006}
{Arthur} S.~J.,  {Hoare} M.~G.,  2006, \mn@doi [\apjs] {10.1086/503899}, \href
  {https://ui.adsabs.harvard.edu/abs/2006ApJS..165..283A} {165, 283}

\bibitem[\protect\citeauthoryear{{Bally} et~al.,}{{Bally}
  et~al.}{2010}]{bally_2010}
{Bally} J.,  et~al., 2010, \mn@doi [\apj] {10.1088/0004-637X/721/1/137}, \href
  {http://ukads.nottingham.ac.uk/abs/2010ApJ...721..137B} {721, 137}

\bibitem[\protect\citeauthoryear{{Bally} et~al.,}{{Bally}
  et~al.}{2014}]{bally_2014}
{Bally} J.,  et~al., 2014, \mn@doi [\apj] {10.1088/0004-637X/795/1/28}, \href
  {http://adsabs.harvard.edu/abs/2014ApJ...795...28B} {795, 28}

\bibitem[\protect\citeauthoryear{{Baranov}, {Krasnobaev}  \&
  {Kulikovskii}}{{Baranov} et~al.}{1971}]{baranov_1971}
{Baranov} V.~B.,  {Krasnobaev} K.~V.,   {Kulikovskii} A.~G.,  1971, Soviet
  Physics Doklady, \href
  {https://ui.adsabs.harvard.edu/abs/1971SPhD...15..791B} {15, 791}

\bibitem[\protect\citeauthoryear{{Barnes}, {Longmore}, {Battersby}, {Bally},
  {Kruijssen}, {Henshaw}  \& {Walker}}{{Barnes} et~al.}{2017}]{barnes_2017}
{Barnes} A.~T.,  {Longmore} S.~N.,  {Battersby} C.,  {Bally} J.,  {Kruijssen}
  J.~M.~D.,  {Henshaw} J.~D.,   {Walker} D.~L.,  2017, \mn@doi [\mnras]
  {10.1093/mnras/stx941}, \href
  {http://adsabs.harvard.edu/abs/2017MNRAS.469.2263B} {469, 2263}

\bibitem[\protect\citeauthoryear{{Barnes}, {Longmore}, {Dale}, {Krumholz},
  {Kruijssen}  \& {Bigiel}}{{Barnes} et~al.}{2020}]{Barnes2020}
{Barnes} A.~T.,  {Longmore} S.~N.,  {Dale} J.~E.,  {Krumholz} M.~R.,
  {Kruijssen} J.~M.~D.,   {Bigiel} F.,  2020, \mn@doi [\mnras]
  {10.1093/mnras/staa2719}, \href
  {https://ui.adsabs.harvard.edu/abs/2020MNRAS.498.4906B} {498, 4906}

\bibitem[\protect\citeauthoryear{{Battersby} et~al.,}{{Battersby}
  et~al.}{2011}]{battersby_2011}
{Battersby} C.,  et~al., 2011, \mn@doi [\aap] {10.1051/0004-6361/201116559},
  \href {http://ukads.nottingham.ac.uk/abs/2011A%26A...535A.128B} {535, A128}

\bibitem[\protect\citeauthoryear{{Battersby} et~al.,}{{Battersby}
  et~al.}{2020}]{Battersby2020}
{Battersby} C.,  et~al., 2020, \mn@doi [\apjs] {10.3847/1538-4365/aba18e},
  \href {https://ui.adsabs.harvard.edu/abs/2020ApJS..249...35B} {249, 35}

\bibitem[\protect\citeauthoryear{{Bressan}, {Marigo}, {Girardi}, {Salasnich},
  {Dal Cero}, {Rubele}  \& {Nanni}}{{Bressan} et~al.}{2012}]{Bressan2012}
{Bressan} A.,  {Marigo} P.,  {Girardi} L.,  {Salasnich} B.,  {Dal Cero} C.,
  {Rubele} S.,   {Nanni} A.,  2012, \mn@doi [\mnras]
  {10.1111/j.1365-2966.2012.21948.x}, \href
  {https://ui.adsabs.harvard.edu/abs/2012MNRAS.427..127B} {427, 127}

\bibitem[\protect\citeauthoryear{{Butterfield}, {Lang}, {Morris}, {Mills}  \&
  {Ott}}{{Butterfield} et~al.}{2018}]{Butterfield2018}
{Butterfield} N.,  {Lang} C.~C.,  {Morris} M.,  {Mills} E. A.~C.,   {Ott} J.,
  2018, \mn@doi [\apj] {10.3847/1538-4357/aa886e}, \href
  {https://ui.adsabs.harvard.edu/abs/2018ApJ...852...11B} {852, 11}

\bibitem[\protect\citeauthoryear{{Callanan} et~al.,}{{Callanan}
  et~al.}{2021}]{Callanan2021}
{Callanan} D.,  et~al., 2021, \mn@doi [\mnras] {10.1093/mnras/stab1527}, \href
  {https://ui.adsabs.harvard.edu/abs/2021MNRAS.505.4310C} {505, 4310}

\bibitem[\protect\citeauthoryear{{Chen}, {Girardi}, {Bressan}, {Marigo},
  {Barbieri}  \& {Kong}}{{Chen} et~al.}{2014}]{Chen2014}
{Chen} Y.,  {Girardi} L.,  {Bressan} A.,  {Marigo} P.,  {Barbieri} M.,   {Kong}
  X.,  2014, \mn@doi [\mnras] {10.1093/mnras/stu1605}, \href
  {https://ui.adsabs.harvard.edu/abs/2014MNRAS.444.2525C} {444, 2525}

\bibitem[\protect\citeauthoryear{{Chen}, {Bressan}, {Girardi}  \&
  {Marigo}}{{Chen} et~al.}{2015}]{Chen2015}
{Chen} Y.,  {Bressan} A.,  {Girardi} L.,   {Marigo} P.,  2015, in IAU General
  Assembly. p. 2257534

\bibitem[\protect\citeauthoryear{{Chevance} et~al.,}{{Chevance}
  et~al.}{2020}]{Chevance2020}
{Chevance} M.,  et~al., 2020, arXiv e-prints, \href
  {https://ui.adsabs.harvard.edu/abs/2020arXiv201013788C} {p. arXiv:2010.13788}

\bibitem[\protect\citeauthoryear{{Clark}, {Glover}, {Ragan}, {Shetty}  \&
  {Klessen}}{{Clark} et~al.}{2013}]{clark_2013}
{Clark} P.~C.,  {Glover} S.~C.~O.,  {Ragan} S.~E.,  {Shetty} R.,   {Klessen}
  R.~S.,  2013, \mn@doi [\apjl] {10.1088/2041-8205/768/2/L34}, \href
  {http://adsabs.harvard.edu/abs/2013ApJ...768L..34C} {768, L34}

\bibitem[\protect\citeauthoryear{{Clark}, {Patrick}, {Najarro}, {Evans}  \&
  {Lohr}}{{Clark} et~al.}{2021}]{Clark2021}
{Clark} J.~S.,  {Patrick} L.~R.,  {Najarro} F.,  {Evans} C.~J.,   {Lohr} M.,
  2021, \mn@doi [\aap] {10.1051/0004-6361/202039205}, \href
  {https://ui.adsabs.harvard.edu/abs/2021A&A...649A..43C} {649, A43}

\bibitem[\protect\citeauthoryear{{Clarkson}, {Ghez}, {Morris}, {Lu}, {Stolte},
  {McCrady}, {Do}  \& {Yelda}}{{Clarkson} et~al.}{2012}]{clarkson_2012}
{Clarkson} W.~I.,  {Ghez} A.~M.,  {Morris} M.~R.,  {Lu} J.~R.,  {Stolte} A.,
  {McCrady} N.,  {Do} T.,   {Yelda} S.,  2012, \mn@doi [\apj]
  {10.1088/0004-637X/751/2/132}, \href
  {https://ui.adsabs.harvard.edu/abs/2012ApJ...751..132C} {751, 132}

\bibitem[\protect\citeauthoryear{{Crowther}}{{Crowther}}{2007}]{Crowther2007}
{Crowther} P.~A.,  2007, \mn@doi [\araa]
  {10.1146/annurev.astro.45.051806.110615}, \href
  {https://ui.adsabs.harvard.edu/abs/2007ARA&A..45..177C} {45, 177}

\bibitem[\protect\citeauthoryear{{Dale}, {Ngoumou}, {Ercolano}  \&
  {Bonnell}}{{Dale} et~al.}{2013}]{Dale2013}
{Dale} J.~E.,  {Ngoumou} J.,  {Ercolano} B.,   {Bonnell} I.~A.,  2013, \mn@doi
  [\mnras] {10.1093/mnras/stt1822}, \href
  {https://ui.adsabs.harvard.edu/abs/2013MNRAS.436.3430D} {436, 3430}

\bibitem[\protect\citeauthoryear{{Davies}, {Hoare}, {Lumsden}, {Hosokawa},
  {Oudmaijer}, {Urquhart}, {Mottram}  \& {Stead}}{{Davies}
  et~al.}{2011}]{davies_2011}
{Davies} B.,  {Hoare} M.~G.,  {Lumsden} S.~L.,  {Hosokawa} T.,  {Oudmaijer}
  R.~D.,  {Urquhart} J.~S.,  {Mottram} J.~C.,   {Stead} J.,  2011, \mn@doi
  [\mnras] {10.1111/j.1365-2966.2011.19095.x}, \href
  {http://adsabs.harvard.edu/abs/2011MNRAS.416..972D} {416, 972}

\bibitem[\protect\citeauthoryear{{Deharveng}, {Pe{\~n}a}, {Caplan}  \&
  {Costero}}{{Deharveng} et~al.}{2000}]{deharveng_2000}
{Deharveng} L.,  {Pe{\~n}a} M.,  {Caplan} J.,   {Costero} R.,  2000, \mn@doi
  [\mnras] {10.1046/j.1365-8711.2000.03030.x}, \href
  {https://ui.adsabs.harvard.edu/abs/2000MNRAS.311..329D} {311, 329}

\bibitem[\protect\citeauthoryear{{Dong} et~al.,}{{Dong}
  et~al.}{2011}]{dong_2011}
{Dong} H.,  et~al., 2011, \mn@doi [\mnras] {10.1111/j.1365-2966.2011.19013.x},
  \href {https://ui.adsabs.harvard.edu/abs/2011MNRAS.417..114D} {417, 114}

\bibitem[\protect\citeauthoryear{{Draine}}{{Draine}}{2011a}]{draine_2011}
{Draine} B.~T.,  2011a, {Physics of the Interstellar and Intergalactic Medium}

\bibitem[\protect\citeauthoryear{{Draine}}{{Draine}}{2011b}]{Draine2011b}
{Draine} B.~T.,  2011b, \mn@doi [\apj] {10.1088/0004-637X/732/2/100}, \href
  {https://ui.adsabs.harvard.edu/abs/2011ApJ...732..100D} {732, 100}

\bibitem[\protect\citeauthoryear{{Federrath} et~al.,}{{Federrath}
  et~al.}{2016}]{federrath_2016}
{Federrath} C.,  et~al., 2016, \mn@doi [\apj] {10.3847/0004-637X/832/2/143},
  \href {http://adsabs.harvard.edu/abs/2016ApJ...832..143F} {832, 143}

\bibitem[\protect\citeauthoryear{{Feldmeier-Krause}, {Kerzendorf}, {Neumayer},
  {Sch{\"o}del}, {Nogueras-Lara}, {Do}, {de Zeeuw}  \&
  {Kuntschner}}{{Feldmeier-Krause} et~al.}{2017}]{Feldmeier-Krause2017}
{Feldmeier-Krause} A.,  {Kerzendorf} W.,  {Neumayer} N.,  {Sch{\"o}del} R.,
  {Nogueras-Lara} F.,  {Do} T.,  {de Zeeuw} P.~T.,   {Kuntschner} H.,  2017,
  \mn@doi [\mnras] {10.1093/mnras/stw2339}, \href
  {https://ui.adsabs.harvard.edu/abs/2017MNRAS.464..194F} {464, 194}

\bibitem[\protect\citeauthoryear{{Figer}, {Kim}, {Morris}, {Serabyn}, {Rich}
  \& {McLean}}{{Figer} et~al.}{1999}]{Figer1999}
{Figer} D.~F.,  {Kim} S.~S.,  {Morris} M.,  {Serabyn} E.,  {Rich} R.~M.,
  {McLean} I.~S.,  1999, \mn@doi [\apj] {10.1086/307937}, \href
  {https://ui.adsabs.harvard.edu/abs/1999ApJ...525..750F} {525, 750}

\bibitem[\protect\citeauthoryear{{Foster} et~al.,}{{Foster}
  et~al.}{2011}]{foster_2011}
{Foster} J.~B.,  et~al., 2011, \mn@doi [\apjs] {10.1088/0067-0049/197/2/25},
  \href {http://adsabs.harvard.edu/abs/2011ApJS..197...25F} {197, 25}

\bibitem[\protect\citeauthoryear{{Geen}, {Bieri}, {Rosdahl}  \& {de
  Koter}}{{Geen} et~al.}{2021}]{Geen2021}
{Geen} S.,  {Bieri} R.,  {Rosdahl} J.,   {de Koter} A.,  2021, \mn@doi [\mnras]
  {10.1093/mnras/staa3705}, \href
  {https://ui.adsabs.harvard.edu/abs/2021MNRAS.501.1352G} {501, 1352}

\bibitem[\protect\citeauthoryear{{Gentry}, {Krumholz}, {Madau}  \&
  {Lupi}}{{Gentry} et~al.}{2019}]{Gentry19a}
{Gentry} E.~S.,  {Krumholz} M.~R.,  {Madau} P.,   {Lupi} A.,  2019, \mn@doi
  [\mnras] {10.1093/mnras/sty3319}, \href
  {https://ui.adsabs.harvard.edu/\#abs/2019MNRAS.483.3647G} {483, 3647}

\bibitem[\protect\citeauthoryear{{Giannetti} et~al.,}{{Giannetti}
  et~al.}{2017}]{Giannetti2017}
{Giannetti} A.,  et~al., 2017, \mn@doi [\aap] {10.1051/0004-6361/201731728},
  \href {https://ui.adsabs.harvard.edu/abs/2017A&A...606L..12G} {606, L12}

\bibitem[\protect\citeauthoryear{{Ginsburg} \& {Kruijssen}}{{Ginsburg} \&
  {Kruijssen}}{2018}]{ginsburg_2018b}
{Ginsburg} A.,  {Kruijssen} J.~M.~D.,  2018, \mn@doi [\apjl]
  {10.3847/2041-8213/aada89}, \href
  {https://ui.adsabs.harvard.edu/abs/2018ApJ...864L..17G} {864, L17}

\bibitem[\protect\citeauthoryear{{Ginsburg}, {Bressert}, {Bally}  \&
  {Battersby}}{{Ginsburg} et~al.}{2012}]{ginsburg_2012}
{Ginsburg} A.,  {Bressert} E.,  {Bally} J.,   {Battersby} C.,  2012, \mn@doi
  [\apjl] {10.1088/2041-8205/758/2/L29}, \href
  {http://ukads.nottingham.ac.uk/abs/2012ApJ...758L..29G} {758, L29}

\bibitem[\protect\citeauthoryear{{Ginsburg} et~al.,}{{Ginsburg}
  et~al.}{2016}]{ginsburg_2016}
{Ginsburg} A.,  et~al., 2016, \mn@doi [\aap] {10.1051/0004-6361/201526100},
  \href {http://adsabs.harvard.edu/abs/2016A%26A...586A..50G} {586, A50}

\bibitem[\protect\citeauthoryear{{Ginsburg} et~al.,}{{Ginsburg}
  et~al.}{2018}]{ginsburg_2018}
{Ginsburg} A.,  et~al., 2018, \mn@doi [\apj] {10.3847/1538-4357/aaa6d4}, \href
  {http://adsabs.harvard.edu/abs/2018ApJ...853..171G} {853, 171}

\bibitem[\protect\citeauthoryear{{Girardi}}{{Girardi}}{2016}]{Girardi2016}
{Girardi} L.,  2016, \mn@doi [\araa] {10.1146/annurev-astro-081915-023354},
  \href {https://ui.adsabs.harvard.edu/abs/2016ARA&A..54...95G} {54, 95}

\bibitem[\protect\citeauthoryear{{Goss}, {Schwarz}, {van Gorkom}  \&
  {Ekers}}{{Goss} et~al.}{1985}]{Goss_1985}
{Goss} W.~M.,  {Schwarz} U.~J.,  {van Gorkom} J.~H.,   {Ekers} R.~D.,  1985,
  \mn@doi [\mnras] {10.1093/mnras/215.1.69P}, \href
  {https://ui.adsabs.harvard.edu/abs/1985MNRAS.215P..69G} {215, 69P}

\bibitem[\protect\citeauthoryear{{Gravity Collaboration} et~al.,}{{Gravity
  Collaboration} et~al.}{2019}]{gravity_2019}
{Gravity Collaboration} et~al., 2019, \mn@doi [\aap]
  {10.1051/0004-6361/201935656}, \href
  {https://ui.adsabs.harvard.edu/abs/2019A&A...625L..10G} {625, L10}

\bibitem[\protect\citeauthoryear{{Green}, {Mackey}, {Haworth}, {Gvaramadze}  \&
  {Duffy}}{{Green} et~al.}{2019}]{green_2019}
{Green} S.,  {Mackey} J.,  {Haworth} T.~J.,  {Gvaramadze} V.~V.,   {Duffy} P.,
  2019, \mn@doi [\aap] {10.1051/0004-6361/201834832}, \href
  {https://ui.adsabs.harvard.edu/abs/2019A&A...625A...4G} {625, A4}

\bibitem[\protect\citeauthoryear{{G{\"u}del}, {Briggs}, {Montmerle}, {Audard},
  {Rebull}  \& {Skinner}}{{G{\"u}del} et~al.}{2008}]{Gudel2008}
{G{\"u}del} M.,  {Briggs} K.~R.,  {Montmerle} T.,  {Audard} M.,  {Rebull} L.,
  {Skinner} S.~L.,  2008, \mn@doi [Science] {10.1126/science.1149926}, \href
  {https://ui.adsabs.harvard.edu/abs/2008Sci...319..309G} {319, 309}

\bibitem[\protect\citeauthoryear{{Habe} \& {Ohta}}{{Habe} \&
  {Ohta}}{1992}]{Habe1992}
{Habe} A.,  {Ohta} K.,  1992, \pasj, \href
  {https://ui.adsabs.harvard.edu/abs/1992PASJ...44..203H} {44, 203}

\bibitem[\protect\citeauthoryear{{Habibi}, {Stolte}  \& {Harfst}}{{Habibi}
  et~al.}{2014}]{habibi_2014}
{Habibi} M.,  {Stolte} A.,   {Harfst} S.,  2014, \mn@doi [\aap]
  {10.1051/0004-6361/201323030}, \href
  {https://ui.adsabs.harvard.edu/abs/2014A&A...566A...6H} {566, A6}

\bibitem[\protect\citeauthoryear{{Haid}, {Walch}, {Seifried}, {W{\"u}nsch},
  {Dinnbier}  \& {Naab}}{{Haid} et~al.}{2018}]{Haid2018}
{Haid} S.,  {Walch} S.,  {Seifried} D.,  {W{\"u}nsch} R.,  {Dinnbier} F.,
  {Naab} T.,  2018, \mn@doi [\mnras] {10.1093/mnras/sty1315}, \href
  {https://ui.adsabs.harvard.edu/abs/2018MNRAS.478.4799H} {478, 4799}

\bibitem[\protect\citeauthoryear{{Hankins}, {Lau}, {Mills}, {Morris}  \&
  {Herter}}{{Hankins} et~al.}{2019}]{Hankins2019}
{Hankins} M.~J.,  {Lau} R.~M.,  {Mills} E.~A.~C.,  {Morris} M.~R.,   {Herter}
  T.~L.,  2019, \mn@doi [\apj] {10.3847/1538-4357/ab174e}, \href
  {https://ui.adsabs.harvard.edu/abs/2019ApJ...877...22H} {877, 22}

\bibitem[\protect\citeauthoryear{{Harper-Clark} \& {Murray}}{{Harper-Clark} \&
  {Murray}}{2009}]{Harper-Clark2009}
{Harper-Clark} E.,  {Murray} N.,  2009, \mn@doi [\apj]
  {10.1088/0004-637X/693/2/1696}, \href
  {https://ui.adsabs.harvard.edu/abs/2009ApJ...693.1696H} {693, 1696}

\bibitem[\protect\citeauthoryear{{Hatchfield} et~al.,}{{Hatchfield}
  et~al.}{2020}]{Hatchfield2020}
{Hatchfield} H.~P.,  et~al., 2020, \mn@doi [\apjs] {10.3847/1538-4365/abb610},
  \href {https://ui.adsabs.harvard.edu/abs/2020ApJS..251...14H} {251, 14}

\bibitem[\protect\citeauthoryear{{Haworth} et~al.,}{{Haworth}
  et~al.}{2015}]{haworth_2015}
{Haworth} T.~J.,  et~al., 2015, \mn@doi [\mnras] {10.1093/mnras/stv639}, \href
  {http://adsabs.harvard.edu/abs/2015MNRAS.450...10H} {450, 10}

\bibitem[\protect\citeauthoryear{{Henshaw} et~al.,}{{Henshaw}
  et~al.}{2016a}]{henshaw_2016}
{Henshaw} J.~D.,  et~al., 2016a, \mn@doi [\mnras] {10.1093/mnras/stw121}, \href
  {http://adsabs.harvard.edu/abs/2016MNRAS.457.2675H} {457, 2675}

\bibitem[\protect\citeauthoryear{{Henshaw}, {Longmore}  \&
  {Kruijssen}}{{Henshaw} et~al.}{2016b}]{henshaw_2016c}
{Henshaw} J.~D.,  {Longmore} S.~N.,   {Kruijssen} J.~M.~D.,  2016b, \mn@doi
  [\mnras] {10.1093/mnrasl/slw168}, \href
  {http://adsabs.harvard.edu/abs/2016MNRAS.463L.122H} {463, L122}

\bibitem[\protect\citeauthoryear{{Henshaw} et~al.,}{{Henshaw}
  et~al.}{2019}]{henshaw_2019}
{Henshaw} J.~D.,  et~al., 2019, \mn@doi [\mnras] {10.1093/mnras/stz471}, \href
  {https://ui.adsabs.harvard.edu/abs/2019MNRAS.485.2457H} {485, 2457}

\bibitem[\protect\citeauthoryear{{Henshaw} et~al.,}{{Henshaw}
  et~al.}{2020}]{henshaw_2020}
{Henshaw} J.~D.,  et~al., 2020, \mn@doi [Nature Astronomy]
  {10.1038/s41550-020-1126-z}, \href
  {https://ui.adsabs.harvard.edu/abs/2020NatAs...4.1064H} {4, 1064}

\bibitem[\protect\citeauthoryear{{Higuchi}, {Chibueze}, {Habe}, {Takahira}  \&
  {Takano}}{{Higuchi} et~al.}{2014}]{higuchi_2014}
{Higuchi} A.~E.,  {Chibueze} J.~O.,  {Habe} A.,  {Takahira} K.,   {Takano} S.,
  2014, \mn@doi [\aj] {10.1088/0004-6256/147/6/141}, \href
  {http://adsabs.harvard.edu/abs/2014AJ....147..141H} {147, 141}

\bibitem[\protect\citeauthoryear{{Hosek}, {Lu}, {Anderson}, {Ghez}, {Morris}
  \& {Clarkson}}{{Hosek} et~al.}{2015}]{hosek_2015}
{Hosek} Matthew~W. J.,  {Lu} J.~R.,  {Anderson} J.,  {Ghez} A.~M.,  {Morris}
  M.~R.,   {Clarkson} W.~I.,  2015, \mn@doi [\apj]
  {10.1088/0004-637X/813/1/27}, \href
  {https://ui.adsabs.harvard.edu/abs/2015ApJ...813...27H} {813, 27}

\bibitem[\protect\citeauthoryear{{Hosokawa} \& {Inutsuka}}{{Hosokawa} \&
  {Inutsuka}}{2006}]{hosokawa_2006}
{Hosokawa} T.,  {Inutsuka} S.-i.,  2006, \mn@doi [\apj] {10.1086/504789}, \href
  {https://ui.adsabs.harvard.edu/abs/2006ApJ...646..240H} {646, 240}

\bibitem[\protect\citeauthoryear{{Hurley}, {Pols}  \& {Tout}}{{Hurley}
  et~al.}{2000}]{Hurley2000}
{Hurley} J.~R.,  {Pols} O.~R.,   {Tout} C.~A.,  2000, \mn@doi [\mnras]
  {10.1046/j.1365-8711.2000.03426.x}, \href
  {https://ui.adsabs.harvard.edu/abs/2000MNRAS.315..543H} {315, 543}

\bibitem[\protect\citeauthoryear{{Immer}, {Menten}, {Schuller}  \&
  {Lis}}{{Immer} et~al.}{2012}]{immer_2012}
{Immer} K.,  {Menten} K.~M.,  {Schuller} F.,   {Lis} D.~C.,  2012, \mn@doi
  [\aap] {10.1051/0004-6361/201219182}, \href
  {http://adsabs.harvard.edu/abs/2012A%26A...548A.120I} {548, A120}

\bibitem[\protect\citeauthoryear{{Jackson} et~al.,}{{Jackson}
  et~al.}{2013}]{jackson_2013}
{Jackson} J.~M.,  et~al., 2013, \mn@doi [\pasa] {10.1017/pasa.2013.37}, \href
  {http://adsabs.harvard.edu/abs/2013PASA...30...57J} {30, e057}

\bibitem[\protect\citeauthoryear{{Jeffreson}, {Kruijssen}, {Krumholz}  \&
  {Longmore}}{{Jeffreson} et~al.}{2018}]{jeffreson_2018}
{Jeffreson} S.~M.~R.,  {Kruijssen} J.~M.~D.,  {Krumholz} M.~R.,   {Longmore}
  S.~N.,  2018, \mn@doi [\mnras] {10.1093/mnras/sty1154}, \href
  {http://adsabs.harvard.edu/abs/2018MNRAS.478.3380J} {478, 3380}

\bibitem[\protect\citeauthoryear{{Johnston}, {Beuther}, {Linz}, {Schmiedeke},
  {Ragan}  \& {Henning}}{{Johnston} et~al.}{2014}]{johnston_2014}
{Johnston} K.~G.,  {Beuther} H.,  {Linz} H.,  {Schmiedeke} A.,  {Ragan} S.~E.,
   {Henning} T.,  2014, \mn@doi [\aap] {10.1051/0004-6361/201423943}, \href
  {http://adsabs.harvard.edu/abs/2014A%26A...568A..56J} {568, A56}

\bibitem[\protect\citeauthoryear{{Kassim} \& {Frail}}{{Kassim} \&
  {Frail}}{1996}]{kassim_1996}
{Kassim} N.~E.,  {Frail} D.~A.,  1996, \mn@doi [\mnras]
  {10.1093/mnras/283.3.L51}, \href
  {http://adsabs.harvard.edu/abs/1996MNRAS.283L..51K} {283, L51}

\bibitem[\protect\citeauthoryear{{Kauffmann}, {Pillai}  \& {Zhang}}{{Kauffmann}
  et~al.}{2013}]{kauffmann_2013}
{Kauffmann} J.,  {Pillai} T.,   {Zhang} Q.,  2013, \mn@doi [\apjl]
  {10.1088/2041-8205/765/2/L35}, \href
  {http://adsabs.harvard.edu/abs/2013ApJ...765L..35K} {765, L35}

\bibitem[\protect\citeauthoryear{{Kauffmann}, {Pillai}, {Zhang}, {Menten},
  {Goldsmith}, {Lu}  \& {Guzm{\'a}n}}{{Kauffmann}
  et~al.}{2017}]{kauffmann_2017a}
{Kauffmann} J.,  {Pillai} T.,  {Zhang} Q.,  {Menten} K.~M.,  {Goldsmith} P.~F.,
   {Lu} X.,   {Guzm{\'a}n} A.~E.,  2017, \mn@doi [\aap]
  {10.1051/0004-6361/201628088}, \href
  {http://adsabs.harvard.edu/abs/2017A%26A...603A..89K} {603, A89}

\bibitem[\protect\citeauthoryear{{Krieger} et~al.,}{{Krieger}
  et~al.}{2017}]{krieger_2017}
{Krieger} N.,  et~al., 2017, \mn@doi [\apj] {10.3847/1538-4357/aa951c}, \href
  {http://adsabs.harvard.edu/abs/2017ApJ...850...77K} {850, 77}

\bibitem[\protect\citeauthoryear{{Kroupa}}{{Kroupa}}{2001}]{kroupa_2001}
{Kroupa} P.,  2001, \mn@doi [\mnras] {10.1046/j.1365-8711.2001.04022.x}, \href
  {http://adsabs.harvard.edu/abs/2001MNRAS.322..231K} {322, 231}

\bibitem[\protect\citeauthoryear{{Kruijssen}, {Longmore}, {Elmegreen},
  {Murray}, {Bally}, {Testi}  \& {Kennicutt}}{{Kruijssen}
  et~al.}{2014}]{kruijssen_2014b}
{Kruijssen} J.~M.~D.,  {Longmore} S.~N.,  {Elmegreen} B.~G.,  {Murray} N.,
  {Bally} J.,  {Testi} L.,   {Kennicutt} R.~C.,  2014, \mn@doi [\mnras]
  {10.1093/mnras/stu494}, \href
  {http://adsabs.harvard.edu/abs/2014MNRAS.440.3370K} {440, 3370}

\bibitem[\protect\citeauthoryear{{Kruijssen}, {Dale}  \&
  {Longmore}}{{Kruijssen} et~al.}{2015}]{kruijssen_2015}
{Kruijssen} J.~M.~D.,  {Dale} J.~E.,   {Longmore} S.~N.,  2015, \mn@doi
  [\mnras] {10.1093/mnras/stu2526}, \href
  {http://adsabs.harvard.edu/abs/2015MNRAS.447.1059K} {447, 1059}

\bibitem[\protect\citeauthoryear{{Kruijssen} et~al.,}{{Kruijssen}
  et~al.}{2019}]{kruijssen_2019}
{Kruijssen} J.~M.~D.,  et~al., 2019, \mn@doi [\mnras] {10.1093/mnras/stz381},
  \href {https://ui.adsabs.harvard.edu/abs/2019MNRAS.484.5734K} {484, 5734}

\bibitem[\protect\citeauthoryear{{Krumholz}}{{Krumholz}}{2017}]{krumholz_2017b}
{Krumholz} M.~R.,  2017, Star Formation.
World Scientific Series in Astrophysics, World Scientific Publishing, Singapore

\bibitem[\protect\citeauthoryear{{Krumholz} \& {Matzner}}{{Krumholz} \&
  {Matzner}}{2009}]{krumholz_2009}
{Krumholz} M.~R.,  {Matzner} C.~D.,  2009, \mn@doi [\apj]
  {10.1088/0004-637X/703/2/1352}, \href
  {http://adsabs.harvard.edu/abs/2009ApJ...703.1352K} {703, 1352}

\bibitem[\protect\citeauthoryear{{Krumholz} et~al.,}{{Krumholz}
  et~al.}{2014}]{Krumholz2014}
{Krumholz} M.~R.,  et~al., 2014, in {Beuther} H.,  {Klessen} R.~S.,
  {Dullemond} C.~P.,   {Henning} T.,  eds, Protostars and Planets VI. p.~243
  (\mn@eprint {arXiv} {1401.2473}),
  \mn@doi{10.2458/azu\_uapress\_9780816531240-ch011}

\bibitem[\protect\citeauthoryear{{Krumholz}, {Kruijssen}  \&
  {Crocker}}{{Krumholz} et~al.}{2017}]{Krumholz2017}
{Krumholz} M.~R.,  {Kruijssen} J.~M.~D.,   {Crocker} R.~M.,  2017, \mn@doi
  [\mnras] {10.1093/mnras/stw3195}, \href
  {https://ui.adsabs.harvard.edu/abs/2017MNRAS.466.1213K} {466, 1213}

\bibitem[\protect\citeauthoryear{{LaRosa}, {Kassim}, {Lazio}  \&
  {Hyman}}{{LaRosa} et~al.}{2000}]{larosa_2000}
{LaRosa} T.~N.,  {Kassim} N.~E.,  {Lazio} T.~J.~W.,   {Hyman} S.~D.,  2000,
  \mn@doi [\aj] {10.1086/301168}, \href
  {http://adsabs.harvard.edu/abs/2000AJ....119..207L} {119, 207}

\bibitem[\protect\citeauthoryear{{Lancaster}, {Ostriker}, {Kim}  \&
  {Kim}}{{Lancaster} et~al.}{2021a}]{Lancaster2021a}
{Lancaster} L.,  {Ostriker} E.~C.,  {Kim} J.-G.,   {Kim} C.-G.,  2021a, \mn@doi
  [\apj] {10.3847/1538-4357/abf8ab}, \href
  {https://ui.adsabs.harvard.edu/abs/2021ApJ...914...89L} {914, 89}

\bibitem[\protect\citeauthoryear{{Lancaster}, {Ostriker}, {Kim}  \&
  {Kim}}{{Lancaster} et~al.}{2021b}]{Lancaster2021b}
{Lancaster} L.,  {Ostriker} E.~C.,  {Kim} J.-G.,   {Kim} C.-G.,  2021b, \mn@doi
  [\apj] {10.3847/1538-4357/abf8ac}, \href
  {https://ui.adsabs.harvard.edu/abs/2021ApJ...914...90L} {914, 90}

\bibitem[\protect\citeauthoryear{{Lang}, {Goss}  \& {Wood}}{{Lang}
  et~al.}{1997}]{lang_1997}
{Lang} C.~C.,  {Goss} W.~M.,   {Wood} O.~S.,  1997, \mn@doi [\apj]
  {10.1086/303452}, \href
  {https://ui.adsabs.harvard.edu/abs/1997ApJ...474..275L} {474, 275}

\bibitem[\protect\citeauthoryear{{Law}, {Backer}, {Yusef-Zadeh}  \&
  {Maddalena}}{{Law} et~al.}{2009}]{law_2009}
{Law} C.~J.,  {Backer} D.,  {Yusef-Zadeh} F.,   {Maddalena} R.,  2009, \mn@doi
  [\apj] {10.1088/0004-637X/695/2/1070}, \href
  {https://ui.adsabs.harvard.edu/abs/2009ApJ...695.1070L} {695, 1070}

\bibitem[\protect\citeauthoryear{{Lindner} et~al.,}{{Lindner}
  et~al.}{2015}]{Lindner2015}
{Lindner} R.~R.,  et~al., 2015, \mn@doi [\aj] {10.1088/0004-6256/149/4/138},
  \href {https://ui.adsabs.harvard.edu/abs/2015AJ....149..138L} {149, 138}

\bibitem[\protect\citeauthoryear{{Lis} \& {Carlstrom}}{{Lis} \&
  {Carlstrom}}{1994}]{lis_1994b}
{Lis} D.~C.,  {Carlstrom} J.~E.,  1994, \mn@doi [\apj] {10.1086/173882}, \href
  {https://ui.adsabs.harvard.edu/abs/1994ApJ...424..189L} {424, 189}

\bibitem[\protect\citeauthoryear{{Lis} \& {Menten}}{{Lis} \&
  {Menten}}{1998}]{lis_1998b}
{Lis} D.~C.,  {Menten} K.~M.,  1998, \mn@doi [\apj] {10.1086/306366}, \href
  {http://adsabs.harvard.edu/abs/1998ApJ...507..794L} {507, 794}

\bibitem[\protect\citeauthoryear{{Lis}, {Menten}, {Serabyn}  \& {Zylka}}{{Lis}
  et~al.}{1994}]{lis_1994}
{Lis} D.~C.,  {Menten} K.~M.,  {Serabyn} E.,   {Zylka} R.,  1994, \mn@doi
  [\apjl] {10.1086/187230}, \href
  {http://adsabs.harvard.edu/abs/1994ApJ...423L..39L} {423, L39}

\bibitem[\protect\citeauthoryear{{Lis}, {Serabyn}, {Zylka}  \& {Li}}{{Lis}
  et~al.}{2001}]{lis_2001}
{Lis} D.~C.,  {Serabyn} E.,  {Zylka} R.,   {Li} Y.,  2001, \mn@doi [\apj]
  {10.1086/319815}, \href {http://adsabs.harvard.edu/abs/2001ApJ...550..761L}
  {550, 761}

\bibitem[\protect\citeauthoryear{{Longmore} et~al.,}{{Longmore}
  et~al.}{2012}]{longmore_2012}
{Longmore} S.~N.,  et~al., 2012, \mn@doi [\apj] {10.1088/0004-637X/746/2/117},
  \href {http://adsabs.harvard.edu/abs/2012ApJ...746..117L} {746, 117}

\bibitem[\protect\citeauthoryear{{Longmore} et~al.,}{{Longmore}
  et~al.}{2013a}]{longmore_2013}
{Longmore} S.~N.,  et~al., 2013a, \mn@doi [\mnras] {10.1093/mnras/sts376},
  \href {http://adsabs.harvard.edu/abs/2013MNRAS.429..987L} {429, 987}

\bibitem[\protect\citeauthoryear{{Longmore} et~al.,}{{Longmore}
  et~al.}{2013b}]{longmore_2013b}
{Longmore} S.~N.,  et~al., 2013b, \mn@doi [\mnras] {10.1093/mnrasl/slt048},
  \href {http://adsabs.harvard.edu/abs/2013MNRAS.433L..15L} {433, L15}

\bibitem[\protect\citeauthoryear{{Longmore} et~al.,}{{Longmore}
  et~al.}{2014}]{longmore_2014}
{Longmore} S.~N.,  et~al., 2014, in {Beuther} H.,  {Klessen} R.~S.,
  {Dullemond} C.~P.,   {Henning} T.,  eds, Protostars and Planets VI. p.~291
  (\mn@eprint {arXiv} {1401.4175}),
  \mn@doi{10.2458/azu\_uapress\_9780816531240-ch013}

\bibitem[\protect\citeauthoryear{{L{\'o}pez-Calder{\'o}n}, {Bronfman}, {Nyman},
  {Garay}, {de Gregorio-Monsalvo}  \& {Bergman}}{{L{\'o}pez-Calder{\'o}n}
  et~al.}{2016}]{Lopez-Calderon2016}
{L{\'o}pez-Calder{\'o}n} C.,  {Bronfman} L.,  {Nyman} L.-{\r{A}}.,  {Garay} G.,
   {de Gregorio-Monsalvo} I.,   {Bergman} P.,  2016, \mn@doi [\aap]
  {10.1051/0004-6361/201321696}, \href
  {https://ui.adsabs.harvard.edu/abs/2016A&A...595A..88L} {595, A88}

\bibitem[\protect\citeauthoryear{{Lu} et~al.,}{{Lu} et~al.}{2019a}]{Lu2019}
{Lu} X.,  et~al., 2019a, \mn@doi [\apjs] {10.3847/1538-4365/ab4258}, \href
  {https://ui.adsabs.harvard.edu/abs/2019ApJS..244...35L} {244, 35}

\bibitem[\protect\citeauthoryear{{Lu} et~al.,}{{Lu} et~al.}{2019b}]{Lu2019b}
{Lu} X.,  et~al., 2019b, \mn@doi [\apj] {10.3847/1538-4357/ab017d}, \href
  {https://ui.adsabs.harvard.edu/abs/2019ApJ...872..171L} {872, 171}

\bibitem[\protect\citeauthoryear{{Luisi} et~al.,}{{Luisi}
  et~al.}{2021}]{Luisi2021}
{Luisi} M.,  et~al., 2021, \mn@doi [Science Advances] {10.1126/sciadv.abe9511},
  \href {https://ui.adsabs.harvard.edu/abs/2021SciA....7.9511L} {7, eabe9511}

\bibitem[\protect\citeauthoryear{{Mac Low} \& {McCray}}{{Mac Low} \&
  {McCray}}{1988}]{MacLow1988}
{Mac Low} M.-M.,  {McCray} R.,  1988, \mn@doi [\apj] {10.1086/165936}, \href
  {https://ui.adsabs.harvard.edu/abs/1988ApJ...324..776M} {324, 776}

\bibitem[\protect\citeauthoryear{{Mac Low}, {van Buren}, {Wood}  \&
  {Churchwell}}{{Mac Low} et~al.}{1991}]{maclow_1991}
{Mac Low} M.-M.,  {van Buren} D.,  {Wood} D. O.~S.,   {Churchwell} E.,  1991,
  \mn@doi [\apj] {10.1086/169769}, \href
  {https://ui.adsabs.harvard.edu/abs/1991ApJ...369..395M} {369, 395}

\bibitem[\protect\citeauthoryear{{Mackey}, {Gvaramadze}, {Mohamed}  \&
  {Langer}}{{Mackey} et~al.}{2015}]{Mackey2015}
{Mackey} J.,  {Gvaramadze} V.~V.,  {Mohamed} S.,   {Langer} N.,  2015, \mn@doi
  [\aap] {10.1051/0004-6361/201424716}, \href
  {https://ui.adsabs.harvard.edu/abs/2015A&A...573A..10M} {573, A10}

\bibitem[\protect\citeauthoryear{{Marigo} et~al.,}{{Marigo}
  et~al.}{2017}]{Marigo2017}
{Marigo} P.,  et~al., 2017, \mn@doi [\apj] {10.3847/1538-4357/835/1/77}, \href
  {https://ui.adsabs.harvard.edu/abs/2017ApJ...835...77M} {835, 77}

\bibitem[\protect\citeauthoryear{{Marsh}, {Ragan}, {Whitworth}  \&
  {Clark}}{{Marsh} et~al.}{2016}]{marsh_2016}
{Marsh} K.~A.,  {Ragan} S.~E.,  {Whitworth} A.~P.,   {Clark} P.~C.,  2016,
  \mn@doi [\mnras] {10.1093/mnrasl/slw080}, \href
  {http://adsabs.harvard.edu/abs/2016MNRAS.461L..16M} {461, L16}

\bibitem[\protect\citeauthoryear{{Mart{\'\i}n-Pintado}, {Gaume},
  {Rodr{\'\i}guez-Fern{\'a}ndez}, {de Vicente}  \&
  {Wilson}}{{Mart{\'\i}n-Pintado} et~al.}{1999}]{Martin-Pintado1999}
{Mart{\'\i}n-Pintado} J.,  {Gaume} R.~A.,  {Rodr{\'\i}guez-Fern{\'a}ndez} N.,
  {de Vicente} P.,   {Wilson} T.~L.,  1999, \mn@doi [\apj] {10.1086/307399},
  \href {https://ui.adsabs.harvard.edu/abs/1999ApJ...519..667M} {519, 667}

\bibitem[\protect\citeauthoryear{{Mart{\'\i}nez-Gonz{\'a}lez}, {Silich}  \&
  {Tenorio-Tagle}}{{Mart{\'\i}nez-Gonz{\'a}lez}
  et~al.}{2014}]{Martinez-Gonzalez2014}
{Mart{\'\i}nez-Gonz{\'a}lez} S.,  {Silich} S.,   {Tenorio-Tagle} G.,  2014,
  \mn@doi [\apj] {10.1088/0004-637X/785/2/164}, \href
  {https://ui.adsabs.harvard.edu/abs/2014ApJ...785..164M} {785, 164}

\bibitem[\protect\citeauthoryear{{Martins}, {Schaerer}  \& {Hillier}}{{Martins}
  et~al.}{2005}]{Martins2005}
{Martins} F.,  {Schaerer} D.,   {Hillier} D.~J.,  2005, \mn@doi [\aap]
  {10.1051/0004-6361:20042386}, \href
  {https://ui.adsabs.harvard.edu/abs/2005A&A...436.1049M} {436, 1049}

\bibitem[\protect\citeauthoryear{{Martins}, {Hillier}, {Paumard}, {Eisenhauer},
  {Ott}  \& {Genzel}}{{Martins} et~al.}{2008}]{martins_2008}
{Martins} F.,  {Hillier} D.~J.,  {Paumard} T.,  {Eisenhauer} F.,  {Ott} T.,
  {Genzel} R.,  2008, \mn@doi [\aap] {10.1051/0004-6361:20078469}, \href
  {https://ui.adsabs.harvard.edu/abs/2008A&A...478..219M} {478, 219}

\bibitem[\protect\citeauthoryear{{Mauerhan}, {Muno}, {Morris}, {Stolovy}  \&
  {Cotera}}{{Mauerhan} et~al.}{2010}]{mauerhan_2010}
{Mauerhan} J.~C.,  {Muno} M.~P.,  {Morris} M.~R.,  {Stolovy} S.~R.,   {Cotera}
  A.,  2010, \mn@doi [\apj] {10.1088/0004-637X/710/1/706}, \href
  {https://ui.adsabs.harvard.edu/abs/2010ApJ...710..706M} {710, 706}

\bibitem[\protect\citeauthoryear{{McKee}, {van Buren}  \& {Lazareff}}{{McKee}
  et~al.}{1984}]{McKee1984}
{McKee} C.~F.,  {van Buren} D.,   {Lazareff} B.,  1984, \mn@doi [\apjl]
  {10.1086/184237}, \href
  {https://ui.adsabs.harvard.edu/abs/1984ApJ...278L.115M} {278, L115}

\bibitem[\protect\citeauthoryear{{McLeod}, {Dale}, {Evans}, {Ginsburg},
  {Kruijssen}, {Pellegrini}, {Ramsay}  \& {Testi}}{{McLeod}
  et~al.}{2019}]{mcleod_2019}
{McLeod} A.~F.,  {Dale} J.~E.,  {Evans} C.~J.,  {Ginsburg} A.,  {Kruijssen}
  J.~M.~D.,  {Pellegrini} E.~W.,  {Ramsay} S.~K.,   {Testi} L.,  2019, \mn@doi
  [\mnras] {10.1093/mnras/sty2696}, \href
  {https://ui.adsabs.harvard.edu/abs/2019MNRAS.486.5263M} {486, 5263}

\bibitem[\protect\citeauthoryear{{Mehringer} \& {Menten}}{{Mehringer} \&
  {Menten}}{1997}]{mehringer_1997}
{Mehringer} D.~M.,  {Menten} K.~M.,  1997, \mn@doi [\apj] {10.1086/303454},
  \href {https://ui.adsabs.harvard.edu/abs/1997ApJ...474..346M} {474, 346}

\bibitem[\protect\citeauthoryear{{Menten}}{{Menten}}{1991}]{menten_1991}
{Menten} K.~M.,  1991, in {Haschick} A.~D.,  {Ho} P. T.~P.,  eds,  Astronomical
  Society of the Pacific Conference Series Vol. 16, Atoms, Ions and Molecules:
  New Results in Spectral Line Astrophysics. pp 119--136

\bibitem[\protect\citeauthoryear{{Mezger} \& {Henderson}}{{Mezger} \&
  {Henderson}}{1967}]{mezger_1967}
{Mezger} P.~G.,  {Henderson} A.~P.,  1967, \mn@doi [\apj] {10.1086/149030},
  \href {http://adsabs.harvard.edu/abs/1967ApJ...147..471M} {147, 471}

\bibitem[\protect\citeauthoryear{{Mezger}, {Pankonin}, {Schmid-Burgk}, {Thum}
  \& {Wink}}{{Mezger} et~al.}{1979}]{Mezger1979}
{Mezger} P.~G.,  {Pankonin} V.,  {Schmid-Burgk} J.,  {Thum} C.,   {Wink} J.,
  1979, \aap, \href {https://ui.adsabs.harvard.edu/abs/1979A&A....80L...3M}
  {80, L3}

\bibitem[\protect\citeauthoryear{{Mills}, {Morris}, {Lang}, {Dong}, {Wang},
  {Cotera}  \& {Stolovy}}{{Mills} et~al.}{2011}]{Mills2011}
{Mills} E.,  {Morris} M.~R.,  {Lang} C.~C.,  {Dong} H.,  {Wang} Q.~D.,
  {Cotera} A.,   {Stolovy} S.~R.,  2011, \mn@doi [\apj]
  {10.1088/0004-637X/735/2/84}, \href
  {https://ui.adsabs.harvard.edu/abs/2011ApJ...735...84M} {735, 84}

\bibitem[\protect\citeauthoryear{{Mills}, {Butterfield}, {Ludovici}, {Lang},
  {Ott}, {Morris}  \& {Schmitz}}{{Mills} et~al.}{2015}]{mills_2015}
{Mills} E.~A.~C.,  {Butterfield} N.,  {Ludovici} D.~A.,  {Lang} C.~C.,  {Ott}
  J.,  {Morris} M.~R.,   {Schmitz} S.,  2015, \mn@doi [\apj]
  {10.1088/0004-637X/805/1/72}, \href
  {http://ukads.nottingham.ac.uk/abs/2015ApJ...805...72M} {805, 72}

\bibitem[\protect\citeauthoryear{{Mills}, {Ginsburg}, {Immer}, {Barnes},
  {Wiesenfeld}, {Faure}, {Morris}  \& {Requena-Torres}}{{Mills}
  et~al.}{2018}]{mills_2018}
{Mills} E.~A.~C.,  {Ginsburg} A.,  {Immer} K.,  {Barnes} J.~M.,  {Wiesenfeld}
  L.,  {Faure} A.,  {Morris} M.~R.,   {Requena-Torres} M.~A.,  2018, \mn@doi
  [\apj] {10.3847/1538-4357/aae581}, \href
  {https://ui.adsabs.harvard.edu/abs/2018ApJ...868....7M} {868, 7}

\bibitem[\protect\citeauthoryear{{Mokiem} et~al.,}{{Mokiem}
  et~al.}{2007}]{Mokiem2007}
{Mokiem} M.~R.,  et~al., 2007, \mn@doi [\aap] {10.1051/0004-6361:20077545},
  \href {https://ui.adsabs.harvard.edu/abs/2007A&A...473..603M} {473, 603}

\bibitem[\protect\citeauthoryear{{Molinari} et~al.,}{{Molinari}
  et~al.}{2011}]{molinari_2011}
{Molinari} S.,  et~al., 2011, \mn@doi [\apjl] {10.1088/2041-8205/735/2/L33},
  \href {http://adsabs.harvard.edu/abs/2011ApJ...735L..33M} {735, L33}

\bibitem[\protect\citeauthoryear{{Morris} \& {Yusef-Zadeh}}{{Morris} \&
  {Yusef-Zadeh}}{1989}]{Morris1989}
{Morris} M.,  {Yusef-Zadeh} F.,  1989, \mn@doi [\apj] {10.1086/167742}, \href
  {https://ui.adsabs.harvard.edu/abs/1989ApJ...343..703M} {343, 703}

\bibitem[\protect\citeauthoryear{{Muijres}, {Vink}, {de Koter}, {M{\"u}ller}
  \& {Langer}}{{Muijres} et~al.}{2012}]{muijres_2012}
{Muijres} L.~E.,  {Vink} J.~S.,  {de Koter} A.,  {M{\"u}ller} P.~E.,   {Langer}
  N.,  2012, \mn@doi [\aap] {10.1051/0004-6361/201015818}, \href
  {https://ui.adsabs.harvard.edu/abs/2012A&A...537A..37M} {537, A37}

\bibitem[\protect\citeauthoryear{{Najarro}, {Figer}, {Hillier}  \&
  {Kudritzki}}{{Najarro} et~al.}{2004}]{najarro_2004}
{Najarro} F.,  {Figer} D.~F.,  {Hillier} D.~J.,   {Kudritzki} R.~P.,  2004,
  \mn@doi [\apjl] {10.1086/423955}, \href
  {https://ui.adsabs.harvard.edu/abs/2004ApJ...611L.105N} {611, L105}

\bibitem[\protect\citeauthoryear{{Newville}, {Stensitzki}, {Allen}  \&
  {Ingargiola}}{{Newville} et~al.}{2014}]{Newville2014}
{Newville} M.,  {Stensitzki} T.,  {Allen} D.~B.,   {Ingargiola} A.,  2014,
  {LMFIT: Non-Linear Least-Square Minimization and Curve-Fitting for Python},
  \mn@doi{10.5281/zenodo.11813}

\bibitem[\protect\citeauthoryear{{Nogueras-Lara} et~al.,}{{Nogueras-Lara}
  et~al.}{2018}]{Nogueras-Lara2018}
{Nogueras-Lara} F.,  et~al., 2018, \mn@doi [\aap]
  {10.1051/0004-6361/201732002}, \href
  {https://ui.adsabs.harvard.edu/abs/2018A&A...610A..83N} {610, A83}

\bibitem[\protect\citeauthoryear{{Nogueras-Lara} et~al.,}{{Nogueras-Lara}
  et~al.}{2019}]{Nogueras-Lara2019}
{Nogueras-Lara} F.,  et~al., 2019, \mn@doi [\aap]
  {10.1051/0004-6361/201936263}, \href
  {https://ui.adsabs.harvard.edu/abs/2019A&A...631A..20N} {631, A20}

\bibitem[\protect\citeauthoryear{{Nogueras-Lara}, {Sch{\"o}del}, {Neumayer},
  {Gallego-Cano}, {Shahzamanian}, {Gallego-Calvente}  \&
  {Najarro}}{{Nogueras-Lara} et~al.}{2020}]{Nogueras-Lara2020}
{Nogueras-Lara} F.,  {Sch{\"o}del} R.,  {Neumayer} N.,  {Gallego-Cano} E.,
  {Shahzamanian} B.,  {Gallego-Calvente} A.~T.,   {Najarro} F.,  2020, \mn@doi
  [\aap] {10.1051/0004-6361/202038606}, \href
  {https://ui.adsabs.harvard.edu/abs/2020A&A...641A.141N} {641, A141}

\bibitem[\protect\citeauthoryear{{Nogueras-Lara}, {Sch{\"o}del}, {Neumayer}  \&
  {Schultheis}}{{Nogueras-Lara} et~al.}{2021a}]{Nogueras-Lara2021a}
{Nogueras-Lara} F.,  {Sch{\"o}del} R.,  {Neumayer} N.,   {Schultheis} M.,
  2021a, \mn@doi [\aap] {10.1051/0004-6361/202140554}, \href
  {https://ui.adsabs.harvard.edu/abs/2021A&A...647L...6N} {647, L6}

\bibitem[\protect\citeauthoryear{{Nogueras-Lara}, {Sch{\"o}del}  \&
  {Neumayer}}{{Nogueras-Lara} et~al.}{2021b}]{Nogueras-Lara2021}
{Nogueras-Lara} F.,  {Sch{\"o}del} R.,   {Neumayer} N.,  2021b, \mn@doi [\aap]
  {10.1051/0004-6361/202040073}, \href
  {https://ui.adsabs.harvard.edu/abs/2021A&A...653A..33N} {653, A33}

\bibitem[\protect\citeauthoryear{{Offner} \& {Arce}}{{Offner} \&
  {Arce}}{2015}]{Offner2015}
{Offner} S. S.~R.,  {Arce} H.~G.,  2015, \mn@doi [\apj]
  {10.1088/0004-637X/811/2/146}, \href
  {https://ui.adsabs.harvard.edu/abs/2015ApJ...811..146O} {811, 146}

\bibitem[\protect\citeauthoryear{{Oka}, {Hasegawa}, {Sato}, {Tsuboi}  \&
  {Miyazaki}}{{Oka} et~al.}{2001}]{Oka2001}
{Oka} T.,  {Hasegawa} T.,  {Sato} F.,  {Tsuboi} M.,   {Miyazaki} A.,  2001,
  \mn@doi [\pasj] {10.1093/pasj/53.5.787}, \href
  {https://ui.adsabs.harvard.edu/abs/2001PASJ...53..787O} {53, 787}

\bibitem[\protect\citeauthoryear{{Ossenkopf} \& {Henning}}{{Ossenkopf} \&
  {Henning}}{1994}]{ossenkopf_1994}
{Ossenkopf} V.,  {Henning} T.,  1994, \aap, \href
  {http://adsabs.harvard.edu/abs/1994A%26A...291..943O} {291, 943}

\bibitem[\protect\citeauthoryear{{Pabst} et~al.,}{{Pabst}
  et~al.}{2019}]{Pabst2019}
{Pabst} C.,  et~al., 2019, \mn@doi [\nat] {10.1038/s41586-018-0844-1}, \href
  {https://ui.adsabs.harvard.edu/abs/2019Natur.565..618P} {565, 618}

\bibitem[\protect\citeauthoryear{{Pabst} et~al.,}{{Pabst}
  et~al.}{2020}]{Pabst2020}
{Pabst} C.~H.~M.,  et~al., 2020, \mn@doi [\aap] {10.1051/0004-6361/202037560},
  \href {https://ui.adsabs.harvard.edu/abs/2020A&A...639A...2P} {639, A2}

\bibitem[\protect\citeauthoryear{{Panagia}}{{Panagia}}{1973}]{panagia_1973}
{Panagia} N.,  1973, \mn@doi [\aj] {10.1086/111498}, \href
  {http://adsabs.harvard.edu/abs/1973AJ.....78..929P} {78, 929}

\bibitem[\protect\citeauthoryear{{Pastorelli} et~al.,}{{Pastorelli}
  et~al.}{2019}]{Pastorelli2019}
{Pastorelli} G.,  et~al., 2019, \mn@doi [\mnras] {10.1093/mnras/stz725}, \href
  {https://ui.adsabs.harvard.edu/abs/2019MNRAS.485.5666P} {485, 5666}

\bibitem[\protect\citeauthoryear{{Pastorelli} et~al.,}{{Pastorelli}
  et~al.}{2020}]{Pastorelli2020}
{Pastorelli} G.,  et~al., 2020, \mn@doi [\mnras] {10.1093/mnras/staa2565},
  \href {https://ui.adsabs.harvard.edu/abs/2020MNRAS.498.3283P} {498, 3283}

\bibitem[\protect\citeauthoryear{{Petkova} et~al.,}{{Petkova}
  et~al.}{2021}]{Petkova2021}
{Petkova} M.~A.,  et~al., 2021, arXiv e-prints, \href
  {https://ui.adsabs.harvard.edu/abs/2021arXiv210409558P} {p. arXiv:2104.09558}

\bibitem[\protect\citeauthoryear{{Pillai}, {Kauffmann}, {Tan}, {Goldsmith},
  {Carey}  \& {Menten}}{{Pillai} et~al.}{2015}]{Pillai2015}
{Pillai} T.,  {Kauffmann} J.,  {Tan} J.~C.,  {Goldsmith} P.~F.,  {Carey} S.~J.,
    {Menten} K.~M.,  2015, \mn@doi [\apj] {10.1088/0004-637X/799/1/74}, \href
  {https://ui.adsabs.harvard.edu/abs/2015ApJ...799...74P} {799, 74}

\bibitem[\protect\citeauthoryear{{Ponti} et~al.,}{{Ponti}
  et~al.}{2015}]{ponti_2015}
{Ponti} G.,  et~al., 2015, \mn@doi [\mnras] {10.1093/mnras/stv1331}, \href
  {http://adsabs.harvard.edu/abs/2015MNRAS.453..172P} {453, 172}

\bibitem[\protect\citeauthoryear{{Portegies Zwart}, {McMillan}  \&
  {Gieles}}{{Portegies Zwart} et~al.}{2010}]{portegies-zwart_2010}
{Portegies Zwart} S.~F.,  {McMillan} S. L.~W.,   {Gieles} M.,  2010, \mn@doi
  [\araa] {10.1146/annurev-astro-081309-130834}, \href
  {https://ui.adsabs.harvard.edu/abs/2010ARA&A..48..431P} {48, 431}

\bibitem[\protect\citeauthoryear{{Rathborne} et~al.,}{{Rathborne}
  et~al.}{2014a}]{rathborne_2014a}
{Rathborne} J.~M.,  et~al., 2014a, \mn@doi [\apj]
  {10.1088/0004-637X/786/2/140}, \href
  {http://adsabs.harvard.edu/abs/2014ApJ...786..140R} {786, 140}

\bibitem[\protect\citeauthoryear{{Rathborne} et~al.,}{{Rathborne}
  et~al.}{2014b}]{rathborne_2014}
{Rathborne} J.~M.,  et~al., 2014b, \mn@doi [\apjl]
  {10.1088/2041-8205/795/2/L25}, \href
  {http://adsabs.harvard.edu/abs/2014ApJ...795L..25R} {795, L25}

\bibitem[\protect\citeauthoryear{{Rathborne} et~al.,}{{Rathborne}
  et~al.}{2015}]{rathborne_2015}
{Rathborne} J.~M.,  et~al., 2015, \mn@doi [\apj] {10.1088/0004-637X/802/2/125},
  \href {http://adsabs.harvard.edu/abs/2015ApJ...802..125R} {802, 125}

\bibitem[\protect\citeauthoryear{{Rathjen} et~al.,}{{Rathjen}
  et~al.}{2021}]{Rathjen2021}
{Rathjen} T.-E.,  et~al., 2021, \mn@doi [\mnras] {10.1093/mnras/stab900}, \href
  {https://ui.adsabs.harvard.edu/abs/2021MNRAS.504.1039R} {504, 1039}

\bibitem[\protect\citeauthoryear{{Riener}, {Kainulainen}, {Henshaw}, {Orkisz},
  {Murray}  \& {Beuther}}{{Riener} et~al.}{2019}]{Riener2019}
{Riener} M.,  {Kainulainen} J.,  {Henshaw} J.~D.,  {Orkisz} J.~H.,  {Murray}
  C.~E.,   {Beuther} H.,  2019, \mn@doi [\aap] {10.1051/0004-6361/201935519},
  \href {https://ui.adsabs.harvard.edu/abs/2019A&A...628A..78R} {628, A78}

\bibitem[\protect\citeauthoryear{{Rodr{\'{\i}}guez} \&
  {Zapata}}{{Rodr{\'{\i}}guez} \& {Zapata}}{2013}]{rodriguez_2013}
{Rodr{\'{\i}}guez} L.~F.,  {Zapata} L.~A.,  2013, \mn@doi [\apjl]
  {10.1088/2041-8205/767/1/L13}, \href
  {http://adsabs.harvard.edu/abs/2013ApJ...767L..13R} {767, L13}

\bibitem[\protect\citeauthoryear{{Rosen}, {Lopez}, {Krumholz}  \&
  {Ramirez-Ruiz}}{{Rosen} et~al.}{2014}]{Rosen2014}
{Rosen} A.~L.,  {Lopez} L.~A.,  {Krumholz} M.~R.,   {Ramirez-Ruiz} E.,  2014,
  \mn@doi [\mnras] {10.1093/mnras/stu1037}, \href
  {http://adsabs.harvard.edu/abs/2014MNRAS.442.2701R} {442, 2701}

\bibitem[\protect\citeauthoryear{{Rosen}, {Offner}, {Foley}  \&
  {Lopez}}{{Rosen} et~al.}{2021}]{Rosen2021}
{Rosen} A.~L.,  {Offner} S. S.~R.,  {Foley} M.~J.,   {Lopez} L.~A.,  2021,
  arXiv e-prints, \href {https://ui.adsabs.harvard.edu/abs/2021arXiv210712397R}
  {p. arXiv:2107.12397}

\bibitem[\protect\citeauthoryear{{Rubin}}{{Rubin}}{1968}]{rubin_1968}
{Rubin} R.~H.,  1968, \mn@doi [\apj] {10.1086/149766}, \href
  {https://ui.adsabs.harvard.edu/abs/1968ApJ...154..391R} {154, 391}

\bibitem[\protect\citeauthoryear{{Schultheis}, {Rich}, {Origlia}, {Ryde},
  {Nandakumar}, {Thorsbro}  \& {Neumayer}}{{Schultheis}
  et~al.}{2019}]{Schultheis2019}
{Schultheis} M.,  {Rich} R.~M.,  {Origlia} L.,  {Ryde} N.,  {Nandakumar} G.,
  {Thorsbro} B.,   {Neumayer} N.,  2019, \mn@doi [\aap]
  {10.1051/0004-6361/201935772}, \href
  {https://ui.adsabs.harvard.edu/abs/2019A&A...627A.152S} {627, A152}

\bibitem[\protect\citeauthoryear{{Schultheis} et~al.,}{{Schultheis}
  et~al.}{2021}]{Schultheis2021}
{Schultheis} M.,  et~al., 2021, \mn@doi [\aap] {10.1051/0004-6361/202140499},
  \href {https://ui.adsabs.harvard.edu/abs/2021A&A...650A.191S} {650, A191}

\bibitem[\protect\citeauthoryear{{Simpson}, {Colgan}, {Cotera}, {Kaufman}  \&
  {Stolovy}}{{Simpson} et~al.}{2018}]{Simpson2018}
{Simpson} J.~P.,  {Colgan} S. W.~J.,  {Cotera} A.~S.,  {Kaufman} M.~J.,
  {Stolovy} S.~R.,  2018, \mn@doi [\apjl] {10.3847/2041-8213/aae8e4}, \href
  {https://ui.adsabs.harvard.edu/abs/2018ApJ...867L..13S} {867, L13}

\bibitem[\protect\citeauthoryear{{Simpson}, {Colgan}, {Cotera}, {Kaufman}  \&
  {Stolovy}}{{Simpson} et~al.}{2021}]{Simpson2021}
{Simpson} J.~P.,  {Colgan} S. W.~J.,  {Cotera} A.~S.,  {Kaufman} M.~J.,
  {Stolovy} S.~R.,  2021, \mn@doi [\apj] {10.3847/1538-4357/abe636}, \href
  {https://ui.adsabs.harvard.edu/abs/2021ApJ...910...59S} {910, 59}

\bibitem[\protect\citeauthoryear{{Smith}}{{Smith}}{2014}]{Smith2014}
{Smith} N.,  2014, \mn@doi [\araa] {10.1146/annurev-astro-081913-040025}, \href
  {https://ui.adsabs.harvard.edu/abs/2014ARA&A..52..487S} {52, 487}

\bibitem[\protect\citeauthoryear{{Smith}, {Norris}  \& {Crowther}}{{Smith}
  et~al.}{2002}]{Smith2002}
{Smith} L.~J.,  {Norris} R. P.~F.,   {Crowther} P.~A.,  2002, \mn@doi [\mnras]
  {10.1046/j.1365-8711.2002.06042.x}, \href
  {https://ui.adsabs.harvard.edu/abs/2002MNRAS.337.1309S} {337, 1309}

\bibitem[\protect\citeauthoryear{{Sormani}, {Tress}, {Glover}, {Klessen},
  {Battersby}, {Clark}, {Hatchfield}  \& {Smith}}{{Sormani}
  et~al.}{2020}]{Sormani2020}
{Sormani} M.~C.,  {Tress} R.~G.,  {Glover} S. C.~O.,  {Klessen} R.~S.,
  {Battersby} C.~D.,  {Clark} P.~C.,  {Hatchfield} H.~P.,   {Smith} R.~J.,
  2020, \mn@doi [\mnras] {10.1093/mnras/staa1999}, \href
  {https://ui.adsabs.harvard.edu/abs/2020MNRAS.497.5024S} {497, 5024}

\bibitem[\protect\citeauthoryear{{Spitzer}}{{Spitzer}}{1978}]{spitzer_1978}
{Spitzer} L.,  1978, {Physical processes in the interstellar medium},
  \mn@doi{10.1002/9783527617722.
}

\bibitem[\protect\citeauthoryear{{Storey} \& {Hummer}}{{Storey} \&
  {Hummer}}{1995}]{Storey1995}
{Storey} P.~J.,  {Hummer} D.~G.,  1995, \mn@doi [\mnras]
  {10.1093/mnras/272.1.41}, \href
  {https://ui.adsabs.harvard.edu/abs/1995MNRAS.272...41S} {272, 41}

\bibitem[\protect\citeauthoryear{{Takahira}, {Tasker}  \& {Habe}}{{Takahira}
  et~al.}{2014}]{Takahira2014}
{Takahira} K.,  {Tasker} E.~J.,   {Habe} A.,  2014, \mn@doi [\apj]
  {10.1088/0004-637X/792/1/63}, \href
  {https://ui.adsabs.harvard.edu/abs/2014ApJ...792...63T} {792, 63}

\bibitem[\protect\citeauthoryear{{Tang}, {Bressan}, {Rosenfield}, {Slemer},
  {Marigo}, {Girardi}  \& {Bianchi}}{{Tang} et~al.}{2014}]{Tang2014}
{Tang} J.,  {Bressan} A.,  {Rosenfield} P.,  {Slemer} A.,  {Marigo} P.,
  {Girardi} L.,   {Bianchi} L.,  2014, \mn@doi [\mnras]
  {10.1093/mnras/stu2029}, \href
  {https://ui.adsabs.harvard.edu/abs/2014MNRAS.445.4287T} {445, 4287}

\bibitem[\protect\citeauthoryear{{Tang}, {Wang}  \& {Wilson}}{{Tang}
  et~al.}{2021}]{Tang2021}
{Tang} Y.,  {Wang} Q.~D.,   {Wilson} G.~W.,  2021, \mn@doi [\mnras]
  {10.1093/mnras/staa3230}, \href
  {https://ui.adsabs.harvard.edu/abs/2021MNRAS.505.2377T} {505, 2377}

\bibitem[\protect\citeauthoryear{{Tielens}}{{Tielens}}{2005}]{tielens_2005}
{Tielens} A.~G.~G.~M.,  2005, {The Physics and Chemistry of the Interstellar
  Medium}

\bibitem[\protect\citeauthoryear{{Tress}, {Sormani}, {Glover}, {Klessen},
  {Battersby}, {Clark}, {Hatchfield}  \& {Smith}}{{Tress}
  et~al.}{2020}]{Tress2020}
{Tress} R.~G.,  {Sormani} M.~C.,  {Glover} S. C.~O.,  {Klessen} R.~S.,
  {Battersby} C.~D.,  {Clark} P.~C.,  {Hatchfield} H.~P.,   {Smith} R.~J.,
  2020, \mn@doi [\mnras] {10.1093/mnras/staa3120}, \href
  {https://ui.adsabs.harvard.edu/abs/2020MNRAS.499.4455T} {499, 4455}

\bibitem[\protect\citeauthoryear{{Tsujimoto}, {Oka}, {Takekawa}, {Yamada},
  {Tokuyama}, {Iwata}  \& {Roll}}{{Tsujimoto} et~al.}{2018}]{Tsujimoto2018}
{Tsujimoto} S.,  {Oka} T.,  {Takekawa} S.,  {Yamada} M.,  {Tokuyama} S.,
  {Iwata} Y.,   {Roll} J.~A.,  2018, \mn@doi [\apj] {10.3847/1538-4357/aab36b},
  \href {https://ui.adsabs.harvard.edu/abs/2018ApJ...856...91T} {856, 91}

\bibitem[\protect\citeauthoryear{{Tsujimoto} et~al.,}{{Tsujimoto}
  et~al.}{2021}]{Tsujimoto2021}
{Tsujimoto} S.,  et~al., 2021, \mn@doi [\apj] {10.3847/1538-4357/abe61e}, \href
  {https://ui.adsabs.harvard.edu/abs/2021ApJ...910...61T} {910, 61}

\bibitem[\protect\citeauthoryear{{Urquhart} et~al.,}{{Urquhart}
  et~al.}{2018}]{Urquhart2018}
{Urquhart} J.~S.,  et~al., 2018, \mn@doi [\mnras] {10.1093/mnras/stx2258},
  \href {https://ui.adsabs.harvard.edu/abs/2018MNRAS.473.1059U} {473, 1059}

\bibitem[\protect\citeauthoryear{{Vink}, {de Koter}  \& {Lamers}}{{Vink}
  et~al.}{2001}]{Vink2001}
{Vink} J.~S.,  {de Koter} A.,   {Lamers} H.~J.~G.~L.~M.,  2001, \mn@doi [\aap]
  {10.1051/0004-6361:20010127}, \href
  {https://ui.adsabs.harvard.edu/abs/2001A&A...369..574V} {369, 574}

\bibitem[\protect\citeauthoryear{{Walker}, {Longmore}, {Bastian}, {Kruijssen},
  {Rathborne}, {Jackson}, {Foster}  \& {Contreras}}{{Walker}
  et~al.}{2015}]{walker_2015}
{Walker} D.~L.,  {Longmore} S.~N.,  {Bastian} N.,  {Kruijssen} J.~M.~D.,
  {Rathborne} J.~M.,  {Jackson} J.~M.,  {Foster} J.~B.,   {Contreras} Y.,
  2015, \mn@doi [\mnras] {10.1093/mnras/stv300}, \href
  {http://adsabs.harvard.edu/abs/2015MNRAS.449..715W} {449, 715}

\bibitem[\protect\citeauthoryear{{Walker} et~al.,}{{Walker}
  et~al.}{2021}]{Walker2021}
{Walker} D.~L.,  et~al., 2021, \mn@doi [\mnras] {10.1093/mnras/stab415}, \href
  {https://ui.adsabs.harvard.edu/abs/2021MNRAS.503...77W} {503, 77}

\bibitem[\protect\citeauthoryear{{Weaver}, {McCray}, {Castor}, {Shapiro}  \&
  {Moore}}{{Weaver} et~al.}{1977}]{weaver_1977}
{Weaver} R.,  {McCray} R.,  {Castor} J.,  {Shapiro} P.,   {Moore} R.,  1977,
  \mn@doi [\apj] {10.1086/155692}, \href
  {https://ui.adsabs.harvard.edu/abs/1977ApJ...218..377W} {218, 377}

\bibitem[\protect\citeauthoryear{{Wilson}, {Rohlfs}  \&
  {H{\"u}ttemeister}}{{Wilson} et~al.}{2009}]{Wilson2009}
{Wilson} T.~L.,  {Rohlfs} K.,   {H{\"u}ttemeister} S.,  2009, {Tools of Radio
  Astronomy}, \mn@doi{10.1007/978-3-540-85122-6.
}

\bibitem[\protect\citeauthoryear{{Yeh}, {Verdolini}, {Krumholz}, {Matzner}  \&
  {Tielens}}{{Yeh} et~al.}{2013}]{Yeh2013}
{Yeh} S. C.~C.,  {Verdolini} S.,  {Krumholz} M.~R.,  {Matzner} C.~D.,
  {Tielens} A. G.~G.~M.,  2013, \mn@doi [\apj] {10.1088/0004-637X/769/1/11},
  \href {https://ui.adsabs.harvard.edu/abs/2013ApJ...769...11Y} {769, 11}

\bibitem[\protect\citeauthoryear{{Yusef-Zadeh}}{{Yusef-Zadeh}}{1989}]{Yusef-Zadeh1989}
{Yusef-Zadeh} F.,  1989, in {Morris} M.,  ed., ~ Vol. 136, The Center of the
  Galaxy. p.~243

\bibitem[\protect\citeauthoryear{{Zhao}, {Desai}, {Goss}  \&
  {Yusef-Zadeh}}{{Zhao} et~al.}{1993}]{Zhao_1993}
{Zhao} J.-H.,  {Desai} K.,  {Goss} W.~M.,   {Yusef-Zadeh} F.,  1993, \mn@doi
  [\apj] {10.1086/173385}, \href
  {https://ui.adsabs.harvard.edu/abs/1993ApJ...418..235Z} {418, 235}

\bibitem[\protect\citeauthoryear{{de Wit}, {Testi}, {Palla}, {Vanzi}  \&
  {Zinnecker}}{{de Wit} et~al.}{2004}]{deWit2004}
{de Wit} W.~J.,  {Testi} L.,  {Palla} F.,  {Vanzi} L.,   {Zinnecker} H.,  2004,
  \mn@doi [\aap] {10.1051/0004-6361:20040454}, \href
  {https://ui.adsabs.harvard.edu/abs/2004A&A...425..937D} {425, 937}

\bibitem[\protect\citeauthoryear{{de Wit}, {Testi}, {Palla}  \&
  {Zinnecker}}{{de Wit} et~al.}{2005}]{deWit2005}
{de Wit} W.~J.,  {Testi} L.,  {Palla} F.,   {Zinnecker} H.,  2005, \mn@doi
  [\aap] {10.1051/0004-6361:20042489}, \href
  {https://ui.adsabs.harvard.edu/abs/2005A&A...437..247D} {437, 247}

\makeatother
\end{thebibliography}



\bsp	
\label{lastpage}
\end{document}